\documentclass[twocolumn]{aastex62}

\submitjournal{ApJ}

\shorttitle{Properties of Free Floating Planets ejected through Planet-Planet Scattering}
\shortauthors{Bhaskar and Perets}

\begin{document}

\title{Properties of Free Floating Planets Ejected Through Planet-Planet Scattering}

\correspondingauthor{Hareesh Gautham Bhaskar}
\email{bhareeshg@gmail.com}

\author{Hareesh Gautham Bhaskar}
\affiliation{Technion-Israel Institute of Technology, Haifa, Israel}

\author{Hagai B. Perets}
\affiliation{Technion-Israel Institute of Technology, Haifa, Israel}



\begin{abstract}
Multiple studies have shown that planet-planet scattering plays an important role in the dynamical evolution of planetary systems. For instance, it has been shown that planet-planet scattering can reproduce the eccentricity distribution of exoplanets. It can also contribute to the current census of free floating planets. In this work we run an ensemble of N-body simulations of planetary systems, and record the properties of planets which are ejected from the system. In our simulations we sample a wide range of orbital and physical properties of the planets. We find that in general $40-80\%$ of the planets are ejected from the system depending on the number of planets initially in the system. Most of the planets are ejected over a timescale of $\sim 10^8-10^9$ years. The ejected planets have a mean excess velocity in the range of  2-6 km/sec with respect to the host star. The excess velocities of the planets ejected from the system strongly depends on the semi-major axis of the inner most planet.  We find that irrespective of their initial location in the planetary system, all planets are equally likely to be ejected from the system. { Also,  bound and ejected planets have distinct mass distributions, with bound planets being more massive than ejected planets. } In addition, increasing the radii of the planets reduces the ejection fraction. This is due to the higher rate of collisions among  the planets. The properties of the ejected planets do not strongly depend on the initial spacing between the planets. The timescale over which ejections happen does increase with the initial separation between the planets. We also find that the ejection fraction does not strongly depend on the distance from the host star beyond which the planets are considered unbound. Finally, we compared our results with observed populations of free floating planets. We conclude that on average 5-10 planets should form around each star to reproduce the observations.
\end{abstract}



\section{Introduction} \label{sec:intro}

Planet-Planet scattering plays a vital role in the dynamical evolution of planetary systems. (\cite{rasioDynamicalInstabilitiesFormation1996}; \cite{weidenschillingGravitationalScatteringPossible1996}; \cite{linOriginMassiveEccentric1997}; \cite{juricDynamicalOriginExtrasolar2008}; \cite{chatterjeeDynamicalOutcomesPlanetPlanet2008}). After formation, planets embedded in the protoplanetary disc are stabilized by the eccentricity damping resulting from the interactions between the planets and the gas in the disc. Within a few million years the gas dissipates, and the planets become more susceptible to scattering encounters (e.g., \cite{chatterjeeDynamicalOutcomesPlanetPlanet2008}). These strong encounters can lead to a variety of outcomes: the planets can collide with each other, they can collide with the star, and they can be ejected from the system. These processes reduce the number of planets bound to host star, leaving a few planets bound to the host star (\cite{juricDynamicalOriginExtrasolar2008}). In general, ejections are more likely than mergers and collisions with the host star. Mergers and collisons with the host star are equally likely if the planets are initialized on near coplanar orbits.

Multiple studies have shown that planet-planet scattering can reproduce the observed distribution of exoplanet eccentricities (\cite{juricDynamicalOriginExtrasolar2008}; \cite{chatterjeeDynamicalOutcomesPlanetPlanet2008}; \cite{2011IAUS..276..225C}; \cite{fordOriginsEccentricExtrasolar2008}; \cite{2009ApJ...699L..88R}; \cite{2010ApJ...711..772R}, \cite{2019A&A...629L...7C}). While scattering alone can explain the eccentricity distribution above 0.2, additional mechanisms are need to explain the low eccentricity distribution. It has been also shown that eccentricity excitation caused by planet-planet scattering can aid in the formation of hot Jupiters \citep{beaugeMultiplePlanetScatteringOrigin2011}. In addition, planet-planet scattering can create conditions that allow high-eccentricity migration of Jupiters through secular processes \citep{lu_planet-planet_2024}. Scattering can also trigger the creation of mini-Oort clouds around planetary systems (\cite{2013MNRAS.429L..99R}). Post-main sequence Planet-Planet scattering can aid in the pollution of the atmospheres of white dwarfs with heavy elements (e.g., \cite{veras_smallest_2023,antoniadou_driving_2019}).

Most studies in literature focus on planets which remain bound to the host star after most of the collisions and ejections have occurred. The properties of planets which are ejected from the systems are less well studied. Ejected planets are also interesting because they can explain a subset of the growing census of free floating planets (FFPs). FFPs are mainly being discovered through microlensing (\cite{sumi_unbound_2011,mrozTwoNewFreefloating2019}), radial velocity and direct imaging surveys (\cite{clanton_constraining_2016}). A diverse set of free-floating planets have been already been discovered including terrestrial planets, Neptunes, and Jovians (e.g., \cite{sumiFreefloatingPlanetMass2023}). In addition, direct imaging has discovered binary planets in star forming regions (\cite{miret-roigRichPopulationFreefloating2021}). Multiple channels for the formation of FFPs have been proposed in literature: they could form either through star formation processes (core collapse \citep{padoanStellarInitialMass2002,miret-roigRichPopulationFreefloating2021,portegieszwartOriginEvolutionWide2024}, photo-evaporation of stellar cores \cite{whitworthFormationFreefloatingBrown2004} and stellar cores ejected from star forming regions \cite{reipurthFormationBrownDwarfs2001}) or through planet formation processes followed by ejections (\cite{2024arXiv240705992C}). The planets can be ejected through planet-planet interactions \citep{verasPlanetPlanetScatteringAlone2012,maFreefloatingPlanetsCore2016}, planet-star interactions in binary systems (\cite{colemanPropertiesFreeFloating2024}), and through stellar flys \citep{yu_free-floating_2024}. In this paper we study the population of planets ejected through planet-planet scattering. 

Previously, \cite{verasPlanetPlanetScatteringAlone2012} studied formation of FFPs through planet-planet scattering. Using an ensemble of N-body simulations, the authors derive the fraction of planets ejected from the system. They find that $\sim 30\%$ of planets are ejected in planetary systems with equal mass planets. The ejection fraction can be increased to upto $70\%$ in systems with unequal mass planets. \ Comparing their results with observations, the authors find that scattering alone cannot fully account for the observed population of FFPs. The authors show that number of giant planets that should form in the systems ($\sim 10-20$ planets) to explained the observed FFPs population ( $\sim 2$ planets per star) is too large to be produced by the conventional planet formation models ($\sim 8$ planets). Population synthesis models of planet formation have been also used to deduce the properties of free-floating planets (\cite{maFreefloatingPlanetsCore2016}). They also find that ejections from planetary systems alone cannot explain the microlensing discoveries of FFPs. But it seems likely that a subset of the observed FFPs are ejected from planetary systems, especially the low mass free floating planets.

A detailed study which looks at the dependence of the ejection fraction, excess velocity, and the mass distribution of the ejected planets on the parameters of the host systems is missing in literature. Such a study would help us delineate the relative importance of various formation channels of FFPs. Nancy Grace Roman Telescope \citep{johnson_predictions_2020}, Euclid Telescope and the Rubin Observatory \citep{bachelet_euclid-roman_2022} will soon discover more FFPs, and detailed theoretical modeling will help us better understand the demographics of FFPs. { It should also be noted that most studies in literature limit their analysis to timescales less than $10^8$ years. On the contrary, we find that some of the properties of FFPs, including ejection fractions can vary significantly over a billion year timescale.}

In this paper we run an ensemble of N-body simulations to study how the properties of the FFPs depend on the initial separation between the planets, mass distribution of the planets, semi-major axis of the inner most planets, mass-radius relationships, mass of the host star, and number of planets initially in the system. In the end, we also describe the properties of planets which remain bound to star. The rest of the paper is organized as following. In Section \ref{sec:setup}, we describe the setup of our simulations. The main results of the paper are described in Section \ref{sec:results}. We discuss our results in Section \ref{sec:discuss}, and conclude in Section \ref{sec:conc}. 

\section{Setup}
\label{sec:setup}
In this study we run an ensemble of $N$-body simulations in which we evolve planetary systems with $N$ planets initialized on near circular and near coplanar orbits around their host stars. The eccentricities of the planets are chosen uniformly between 0 and 0.01, and their inclinations are chosen between  0 and $1^\circ$. Other angles are randomly assigned between 0 and $2\pi$. The number of planets  initially in the system ($N$) is chosen between 3 and 10. The planets are labeled using indices starting from 0 for the inner most planet, to $N-1$ for the outermost planet. The labels are assigned at the beginning of the simulation, and are not changed later. The mass, radius, semi-major axis, eccentricity, inclination, argument of pericenter and longitude of pericenter of planet$-i$ are denoted by $m_i,R_i, a_i,e_i,I_i,\omega_i$ and $\Omega_i$ respectively. Also, the mass of the central star is $m_s$ and it's radius is $R_s$. In all of our simulations  we set $m_s=1$ $M_\odot$ and $R_s=1$ $R_\odot$. The semi-major axis of the innermost planet is varied between 1 and 10 AU. We choose different distributions for the masses of the Planets. In our fiducial runs we assume that all the planets are Jupiter mass objects. The initial timestep is chosen to be 1/100th the orbital period of the inner most planet in the system. We use the MERCURIUS integrator in Rebound package to run our simulations \citep{reinREBOUNDOpensourceMultipurpose2012}.

For our fiducial simulation we assume that the radius of the planet is 1 $R_{jup}$. To model collisions, we use the sticky sphere approximation implemented in Rebound. It should be noted that the tidal interactions between the planets can significantly increase the effective radius of the planets (upto a factor of 2). This can affect the results of the simulations (e.g. \cite{liGiantPlanetScatterings2021a}). Hence, we run additional simulations in which we increase the radii of the planets. We also run a set of simulations in which we ignore collisions by setting the radii of the planets to 0.

When the planets reach a distance of $10^5$ AU from the central star, the planets are assumed to be ejected from the system. The ejected planets are removed from the simulation and their velocities with respect to the host star are recorded. Because of observational constraints, the separation between the star and the planets is not always constrained. Hence, in some of the simulations, we also change the distance beyond which the planets are considered ``free floating" (also called ejection distance, $D$). The ejection distance, $D$ is sampled between $100$ AU and $10^5$ AU. For each planet ejected from the system we calculate its excess speed ($v_\infty$) which is given by:
\begin{equation}
    v_\infty = \sqrt{v^2-\frac{Gm_s}{D}}
\end{equation}


It should be noted that the timescale over which planetary systems become unstable as a result of planet-planet scattering depends on the initial separation between the planets. For instance, \cite{chatterjeeDynamicalOutcomesPlanetPlanet2008} derive an empirical relationship between the instability timescale ($t_{instability}$) and the initial separation in mutual hill radii ($K$):  $\log_{10}[t_{instability}/P_{0,init}]=a+b\exp{(cK)}$, where $a=1.07$, $b=0.03$ and $c=1.10$. For $K=4$, this gives us $t_{instability}=3.3\times 10^3$ years for $a=1$ AU and $t_{instability}=1\times 10^5$ years for $a=10$ AU. Hence to properly model the instability triggered by scattering, we end the simulations either when the number of bound planets in the system is 1, or when the time limit of $t_{max}$ is reached. Unless otherwise stated $t_{max}$ is taken to be $10^9$ years.  In some of our simulations we also change the initial mutual separation between the planets. It is sampled between $K=3$ and $5$.

In the following subsections we will discuss at the results of the simulations.
\section{Results}
\label{sec:results}
\subsection{Fiducial Simulations}

In this section we discuss results of our fiducial simulations. In our fiducial simulations, we assume that the planetary system initially has 5 planets. The inner most planet of the system is initialized on an orbit with a semi-major axis of 3 AU. The planets are initially placed at a separation of $K=4$ mutual hill radii from each other. All the planets initially have the same mass (1 $M_{jup}$) and radii (1 $R_{jup}$). The planets are considered to be unbound if they reach a distance of $10^5$ AU from the host star. We evolved 100 such systems in our simulations.

\begin{figure}
	\includegraphics[width=0.8\linewidth]{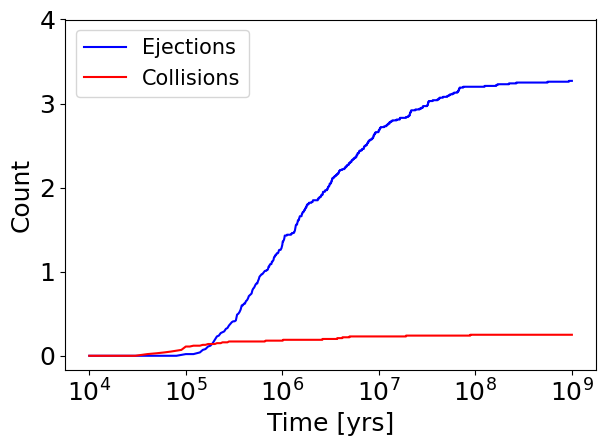}
	\caption{The average number of collisions (red) and ejections (blue) as a function of time in our fiducial simulations. We can see that most of the ejections happen in $10^8$ years. Meanwhile collisions happen on much shorter timescales ($\sim 10^6$ years). On average $\sim 3$ planets are ejected from the system as compared to 0.4 collisions. We use the following parameters to run the simulations: $N_p=5, a_{0,init}=3$ AU, $m_s=1$ $M_\odot,m_j=1$ $M_{jup}$ and $K$=4. We evolve 100 planetary systems in our simulations.}
	\label{fig:fidbr}
\end{figure}

Figure \ref{fig:fidbr} shows the average number of ejections and collisions as a function of time. We can see that around 3.5 planets are ejected from the system in $10^9$ years. Most of the ejections happen within $10^8$ years. Meanwhile, far fewer planets ($\sim 0.4$ collisions on average) undergo collisions. We find that, in general, planets which are initialized on orbits close to the host star are ejected first. This is due to the fact that they have shorter dynamical timescales. We also find that all the planets are equally likely to be ejected from the system irrespective of their initial location in the planetary system (see Figure \ref{fig:hashplanets}).  We also find that all the planets which are ejected from the system have the same mass as the initial mass of the planets. Meanwhile, heavier planets which are the product of planet-planet collisions remain bound to the host star.

\begin{figure}
	\includegraphics[width=1.0\linewidth]{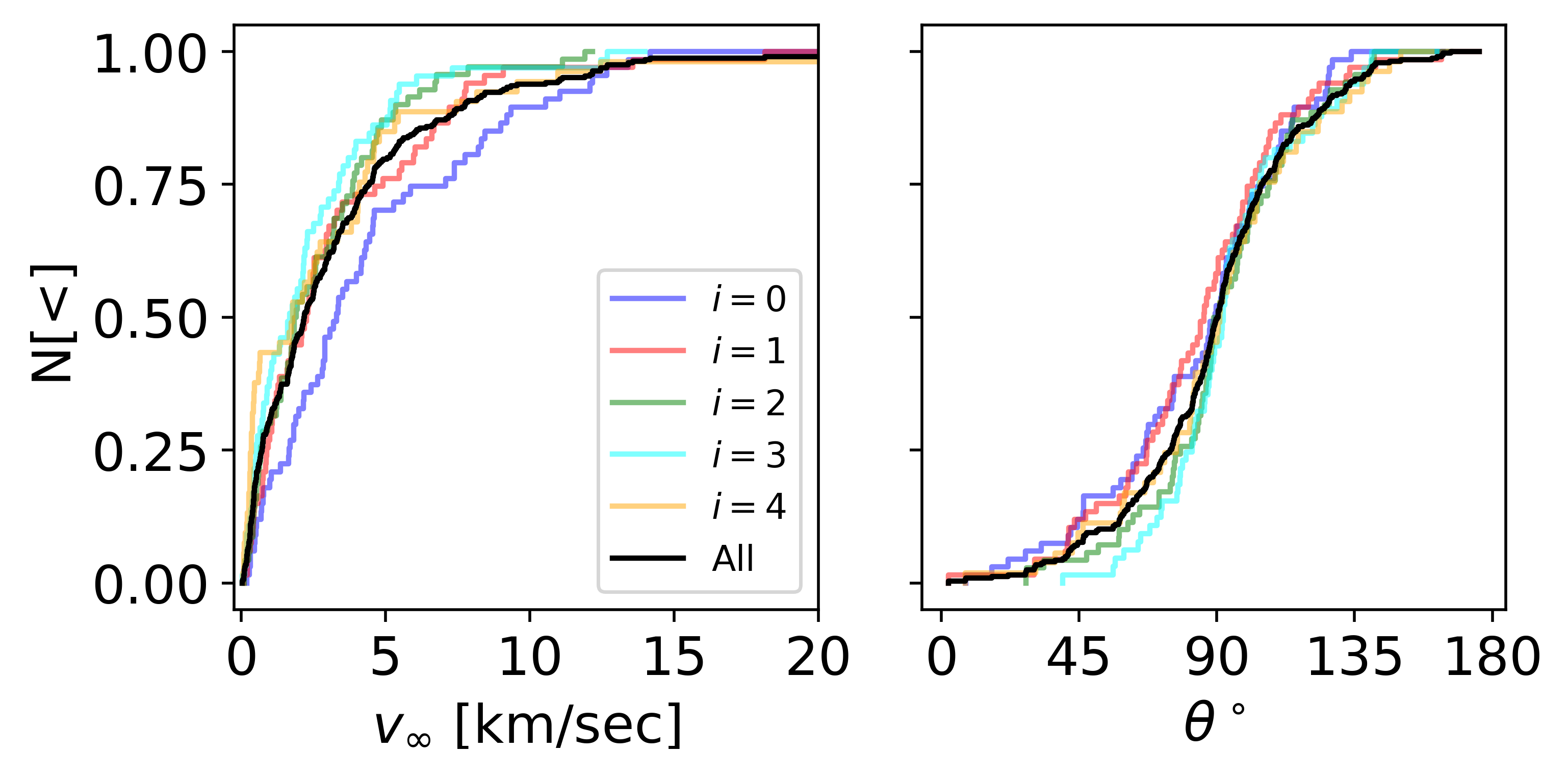}
	\caption{The left panel shows the distribution of excess velocities of ejected planets (with respect to their host stars) in our fiducial simulations. We can see that most of the ejected planets have excess speeds less than a 10 km/sec. We find the mean excess speed to be 2.1 km/sec. The right panel shows the distribution of ejection angle with respect to the initial angular momentum of the planetary system. The results for the entire ensemble is shown using black lines. The distribution for a specific planet index is shown using different colors. We can see that most of the planets are ejected in the plane of the planetary system ($\theta \sim 90^\circ$), with $72\%$ of planets ejected within $30^\circ$ of the initial planetary plane. We use the following parameters to run the simulations: $N_p=5, a_{0,init}=3$ AU, $m_s=1$ $M_\odot,m_j=1$ $M_{jup}$ and $K$=4. We evolve 100 planetary systems in our simulations. }
	\label{fig:fidveldthetad}
\end{figure}

We will now look the excess velocities of planets which were ejected from the planetary system. The left panel of Figure \ref{fig:fidveldthetad} shows the distribution of excess velocities with respect to the host stars. We can see that most of the planets have excess velocities less than 15 km/sec. The median and mean excess velocities were found to be 3.5 and 2.2 km/sec respectively. The excess velocity of the ejected planet depends on it’s initial location in the planetary system, with planets initially closer to the host star ejected with higher velocities. This is shown in Figure \ref{fig:fidveldthetad} using various colors. In our simulations, escaping planets which were initialized closest to the host star had an average excess velocity of 5.1 km/sec. In comparison, ejected planets initially on the widest orbits have an average excess velocity of 2.5 km/sec. These results are consistent with other studies in literature (e.g., \cite{moorheadGiantPlanetMigration2005}).

We also record the direction in which the planets were ejected. The right panel of Figure \ref{fig:fidveldthetad} shows the angle between the velocity of the ejected planet and the initial angular momentum of the planetary system. We can see that most of the planets are ejected in the plane of the planetary system, with around $72\%$ of planets ejected within $30^0$ of the initial plane of the planetary system. Also, planets initialized closer to the host star are more likely to ejected out of the initial plane of the planetary system as compared to planets initially on wider orbits.
\subsection{Dependence on the number of planets in the system}
Multiplanetary systems are very common in the solar neighborhood. Several studies have deduced the intrinsic multiplicity of exosystems by accounting for the observational biases in the exoplanet surveys. For instance, radial velocity surveys have shown that sun-like stars hosts, on average, close to 2 planets (e.g., \cite{zhuIntrinsicMultiplicityDistribution2022}). Meanwhile, transit surveys show that most transiting planets have a companion (e.g, \cite{zinkAccountingIncompletenessDue2019a}). It should be noted that the observed exosystems are products of billions of years of dynamical evolution. As discussed in previous sections, planetary systems are expected to undergo a phase of dynamical instability during which collisions and ejections of the planets are likely. Hence, we expect more planets to be form in planetary systems as compared to what we observe today.


Both core accretion and gravitational instability theories provide estimates for the number of planets that can form in a protoplanetary disc. For instance, core accretion theory can produce 8-10 planets within 35 AU \citep{dodson-robinson_formation_2009}.  Meanwhile gravitational instability can produce a few giant planets at larger radii ($>35$ AU) \citep{boss_formation_2006,boley_two_2009,boley_clumps_2010}. It should be noted that irrespective of their formation location, the planets can also migrate either inward or outward in the disc. The exact number of planets produced in the system depends on the details of the protoplanetary disc. To account for the various uncertainities, we vary the number of planets initially in the system in our simulations.

\begin{figure}
	\centering
	\includegraphics[scale=0.5]{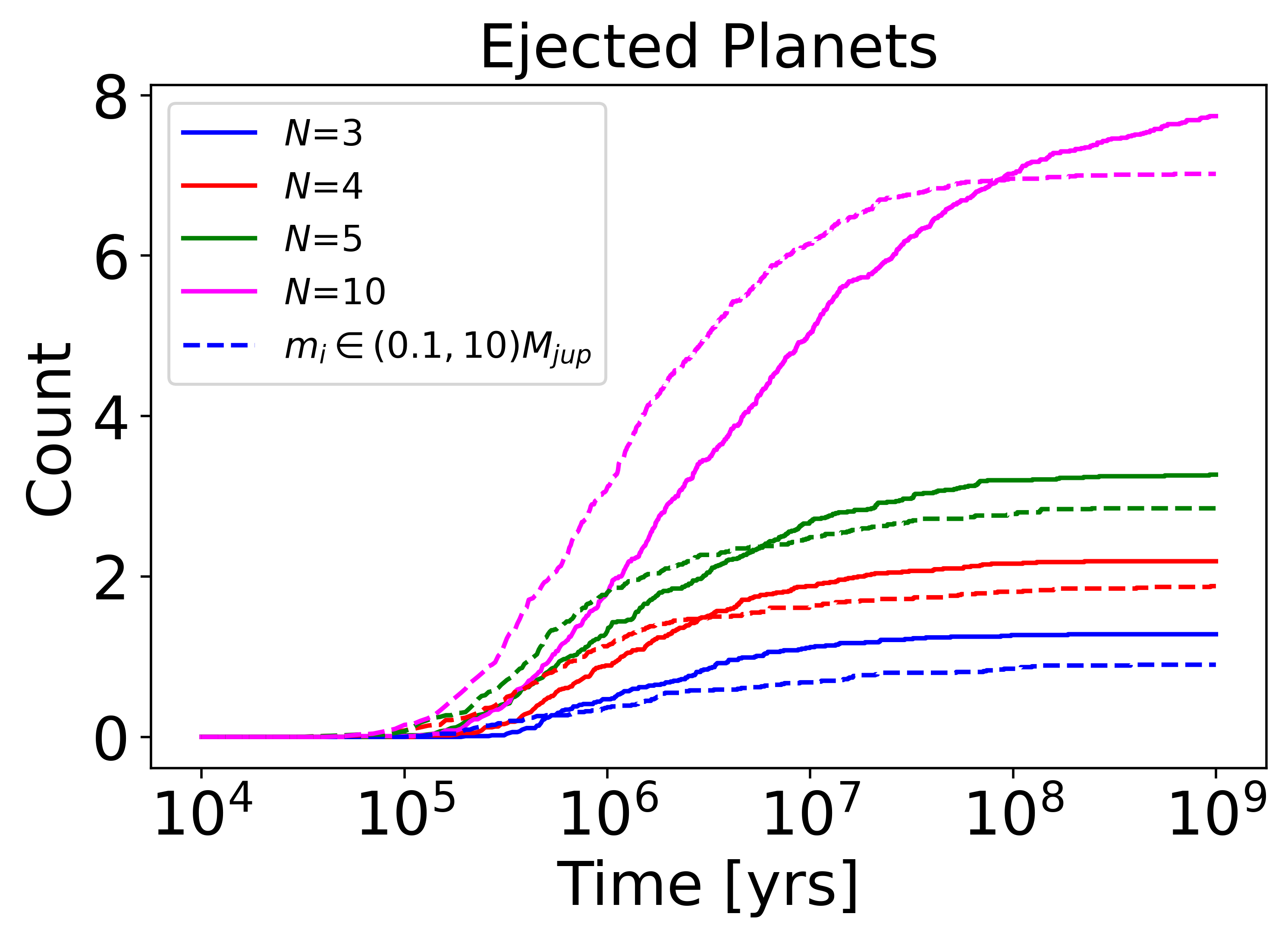}
	\caption{Dependence on the number of planets initially in the system. The average number of planets ejected from the system is shown on the y axis and the time is shown on the x-axis. The colors show the number of planets initially in the system. The solid lines show results for systems in which the planets initially have identical masses. Results shown using dashed line correspond to systems in which the initial planet masses were sampled from a log uniform distribution. For $N=3,4$ and 5, most of the planets are ejected from the systen in $\sim 10^8$ years. Meanwhile, when $N=10$ not all planets are ejected in $10^9$ years. In unequal mass simulations, the ejection fraction reaches the steady state on shorter timescales. Also, fewer planets are ejected from the system in these systems. We use the following parameters to run the simulations: $N_p=5, a_{0,init}=3$ AU, $m_s=1M_\odot,m_j=1M_{jup}$ and $K$=4. We evolve 100 systems in our simulations for each choice of $N$.}
	\label{fig:nej_vs_t_nplanets}
\end{figure}

In this section we change the number of planets initially in the system between 3 and 10. The results are shown in Figure \ref{fig:nej_vs_t_nplanets}. It shows the average number of planets ejected from the system as a function of time. We can see that most of planets are ejected from the system in $10^8$ years when $N=3,4$ and $5$. Meanwhile, for $N=10$ the average number of planets ejected from the system does not reach a steady state even in $10^9$ years. Also the ejection fraction increases with $N$: for $N=3$ the ejection fraction is $45\%$ which increases to $70\%$ for $N=10$. In general, after $10^9$ years of evolution only $1-2$ planets remain bound to the host star irrespective of number of planets initially in the system.

\begin{figure}
	\centering
	\includegraphics[scale=0.5]{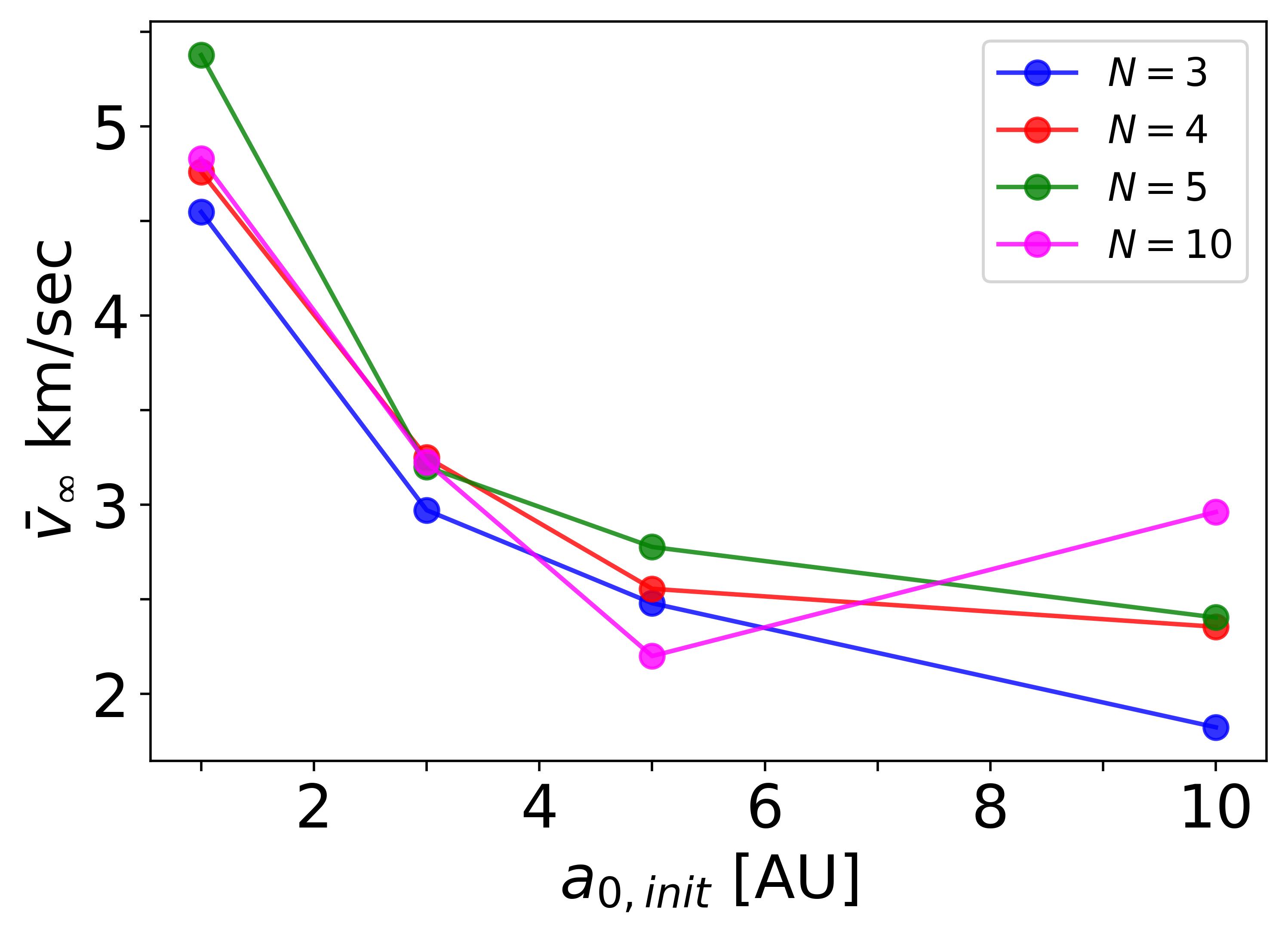}
	\caption{Dependence of the mean excess velocity with respect to the host star (y-axis) on the initial semi-major axis of the inner most planet, $a_{0,init}$ (x-axis). The colors show the number of planets initially in the system. We can see that the excess velocity decreases with the semi-major axis of the innermost planet. Also, the dependence of the excess velocity on the number of planets in the system is weak. We use the following parameters to run the simulations:  $m_s=1 M_\odot,m_j=1M_{jup}$ and $K=4$. We evolve 100 planetary systems for each choice of $a_{0,init}$ and $N$.}
	\label{fig:vescsma}
\end{figure}
 
Figure \ref{fig:vescsma} shows the average excess speed of the ejected planets as a function of the initial semi-major axis of the innermost planet. The colors show the number of planets initially in the system. We can see that the excess speed of the ejected planets does not strongly depend on the number of planets initially in the system. Meanwhile, the average excess velocity decreases with the semi-major axis of the innermost planet (see Section \ref{sec:subsecsma}).
\begin{figure}
	\centering
	\includegraphics[scale=0.5]{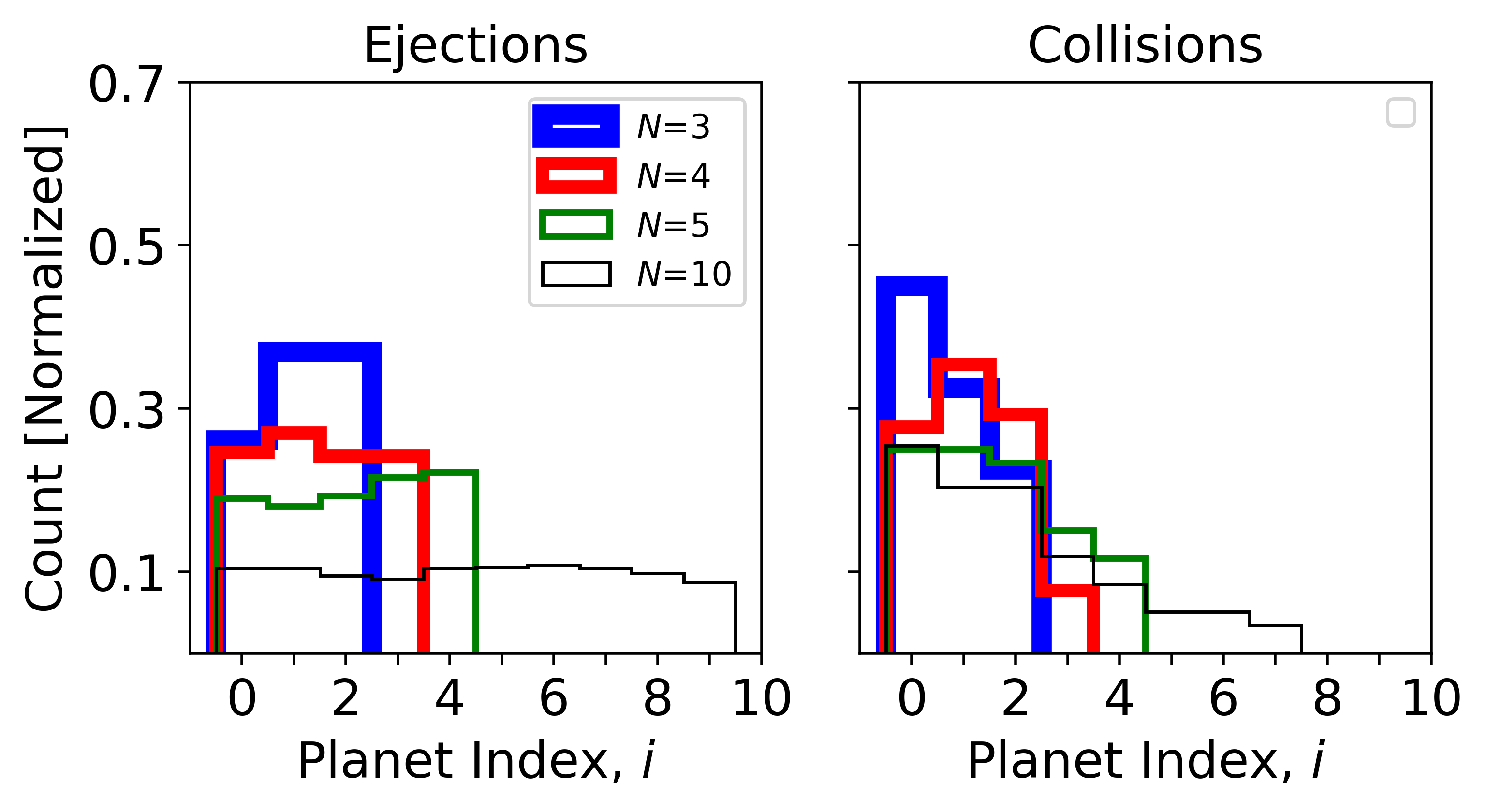}	
	\caption{The distribution of planets which have either been ejected (left panel) or have collided (right) during the course of the simulation. The x-axis shows the planet index, and colors show the initial number of planets in the system. Planets with an index of 0 are initialized closest to their host stars, and those with index $N-1$ are on the widest orbits. We can see that collisions mostly happen in close-in planets, and planets on wide orbits are more likely to be ejected. We use the following parameters to run the simulations: $a_{0,init}=3$ AU, $m_s=1$ $M_\odot,m_j=1$ $M_{jup}$ and $K$=4. We evolve 100 planetary systems for each choice of $N$.}
	\label{fig:hashplanets}
\end{figure}

Figure \ref{fig:hashplanets} shows the distributions of planets which are ejected (left panel) or have undergone collisions (right panel) as a function of planet index. We can see that collisions mostly happen between 3-4 inner planets in the system. This is consistent with the fact that these planets have large orbital velocities, and hence a lower Safronov number which makes collisions more likely (e.g., \cite{tremaineDynamicsPlanetarySystems2023}). Meanwhile, the left panel shows that all planets are equally likely to be ejected from the system irrespective of their initial location. We also find that that planets which are initialized on wider orbits are likely to be ejected first.

\subsection{Dependence on the semi-major axis of the inner-most planet.}
\label{sec:subsecsma}
\begin{figure}
	\centering
	\includegraphics[scale=0.55]{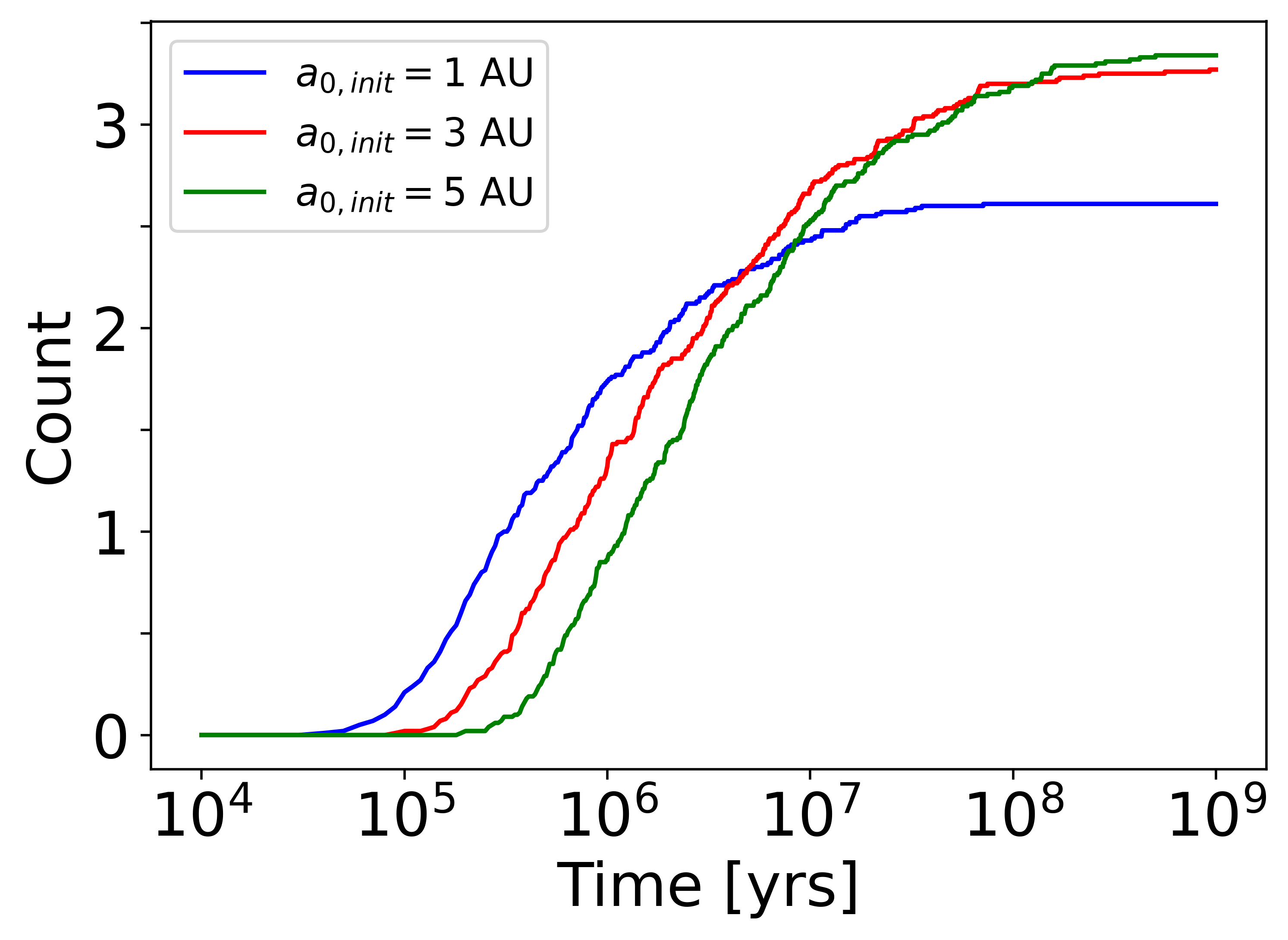}
	\caption{Dependence on the semi-major axis of the inner most planet. The average number of planets ejected from the system as a function of time. The colors show the semi-major axis of the inner-most planets. The ejection fraction increases with the semimajor axis of the inner most planets. This is due to collisions which are dominant in more compact systems. The timescale over which the ejections occur increases with the semi-major axis of the inner most planet. We use the following parameters to run our simulations: $N=5$ , $m_s=1 M_\odot,m_j=1 M_{jup}$ and $K=4$.}
	\label{fig:avgejsma}
\end{figure}
The observed exosystems have diverse architectures. For instance, the separation of the innermost planet in multiplanetary systems from their host stars ranges from a few percent of an AU to 100s of AU. For instance, KOI-55 has two terrestrial ultra short period planets within 0.008 AU. Meanwhile, TYC 8998-760-1 is a young stellar system with two planets beyond 160 AU. We find that in most of the observed exosystems ($\sim 95\%$) \footnote{Based on data from Nasa exoplanet archive.}, the semi-major axis of the innermost planet is within 5 AU, with a mean of 0.45 AU. It should be noted that due to observational biases, the intrinsic distribution of semi-major axes maybe different. But, the current observations do tell us about the diversity of exoplanet architectures. 

Planet formation models also provides constraints on the locations of planets formation. For instance, core accretion models can produce multiple Jupiter mass planets at a few AU around M-type stars. Gravitational instability is more effective farther away from the host star. It is expected to produce a few Jupiter mass planets beyond 35 AU. After formation, these planets are expected to undergo dynamical evolution, including migration due to disc-planet interactions, planet-planet secular interactions and planet-planet scattering. To account for these uncertainties, we run a set of simulations in which we change the semi-major axis of the inner most planet in the planetary system.

The properties of ejected planets can strongly depend on the semi-major axis of the inner most planets. When the inner planets are close to the host star, they are more likely to have collisions with each other. This would reduce the number of planets ejected from the system. Also, the timescale over which the ejections happen is also shorter when the semi-major axis of the inner most planet is short. This is due to the shorter dynamical timescale of the system. This can be seen in Figure \ref{fig:avgejsma} which shows ejection fraction as a function of time. Colors show results from runs with different values for the semi-major axis of the innermost planet. As expected, we can see the increase in the final ejection fraction (at $\sim 10^9$ years) with the semi-major axis of the inner most planet. The timescales over which the ejections happen also increases with the $a_{0,init}$.

\begin{figure}
	\centering
	\includegraphics[scale=0.5]{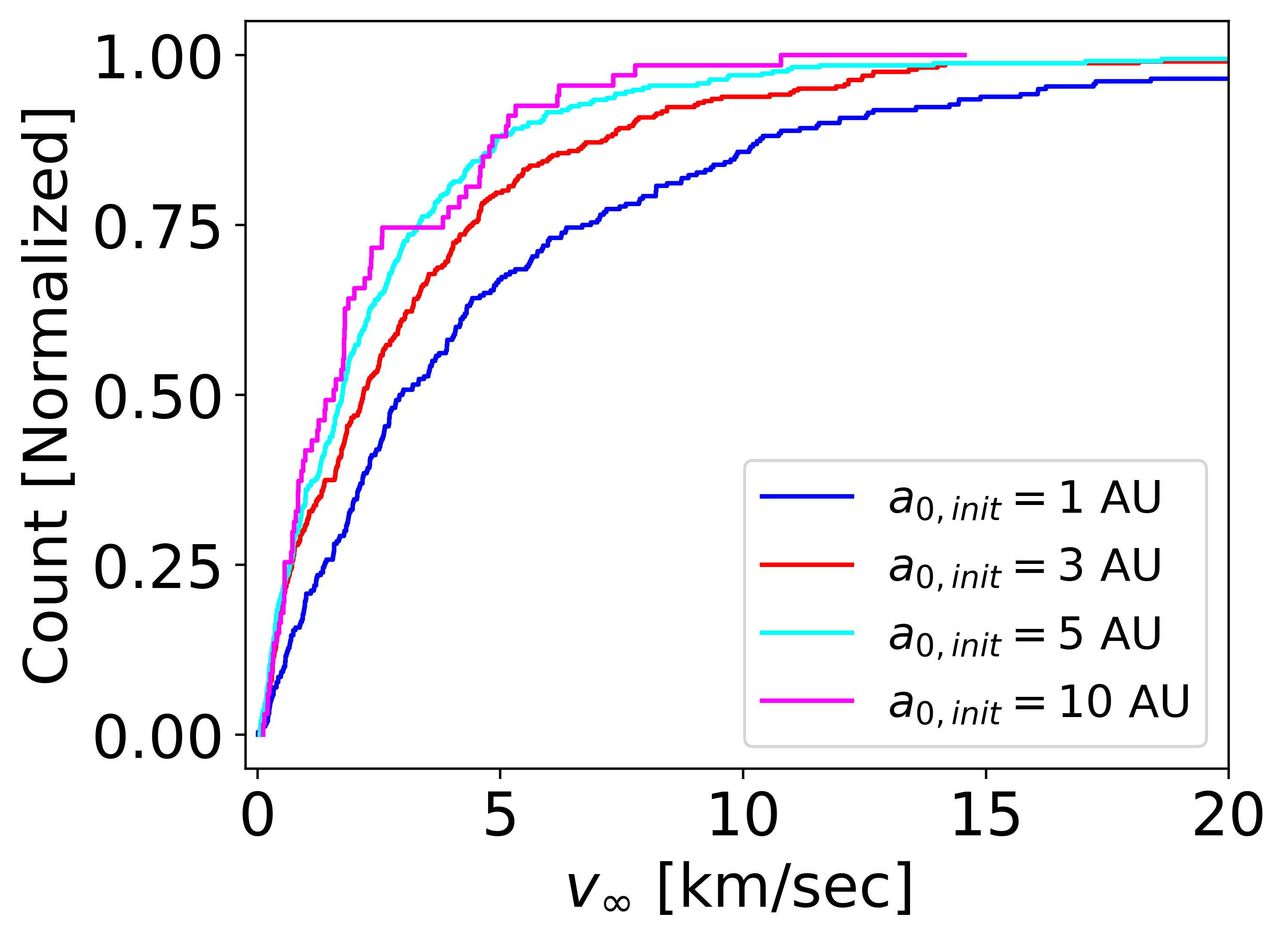}
	\caption{Dependence of excess velocity on the semi-major axis of the inner most planet. The colors show the semi-major axis of the inner-most planets. The excess velocity decreases with the semi-major axis of the inner most planets. We use the following parameters to run the simulations: $N=5$, $m_s=1M_\odot,m_j=1M_{jup}$ and $K=4$.}
	\label{fig:velescsma}
\end{figure}

The excess velocity distribution also depends on the semi-major of the inner most planet. This can be seen in Figure \ref{fig:velescsma}, which shows the cumulative distribution of excess velocities. The colors show the semi-major axes of the inner most planet.  We can see that the distribution peaks at higher velocities as we decrease the semi-major axis of the inner most planet. For instance the velocity distribution peaks at $\sim 3.5$ km/sec for $a_{0,init}=1$ AU, while for $a_{0,init}=5$ AU the velocity distribution peaks at $\sim 2$ km/sec. Similarly, Figure \ref{fig:vescsma} which shows that the mean excess velocity decreases with the semi-major axis of the inner most planet.

\subsection{Dependence on the initial masses of the planets}
In this study so far we have focused on systems in which the masses of all the planets are identical. But, the observed exosystems have a wide range of masses and radii. Hence in this section we sample the masses from a log-uniform distribution between 0.1 $M_{jup}$ and 10 $M_{jup}$. In our simulations we also change the number of planets in system to account for the variable outcomes of planet formation.


The dashed lines in Figure \ref{fig:nej_vs_t_nplanets} show the results of simulations in which the mass of the planets is sampled from a log uniform distribution. At the end of the simulation, we can see that fewer planets are ejected from systems in which the planets initially have an unequal mass distribution. On average, number of planets ejected the system is reduced by 1. Also, the ejection fraction reaches steady state faster in unequal mass systems as compared to systems with identical planets. Consequently, up to a few million years of evolution, more planets are ejected from unequal mass system as compared to systems with equal mass planets. We also find that number of collisions are also lower in systems with unequal mass distribution. Meanwhile, more planets remain bound to the host star in such systems. Compared to planets ejected from systems with identical planets, we find that planets ejected from unequal mass systems tend to have lower ejection speeds. For example, when $N=5$, the mean excess speed for unequal (equal) mass systems is 3.25 km/sec (3.15 km/sec). Also, more of the ejected planets are ejected in the initial plane of the planetary system when the masses are unequal.

\begin{figure}
 \includegraphics[scale=0.55]{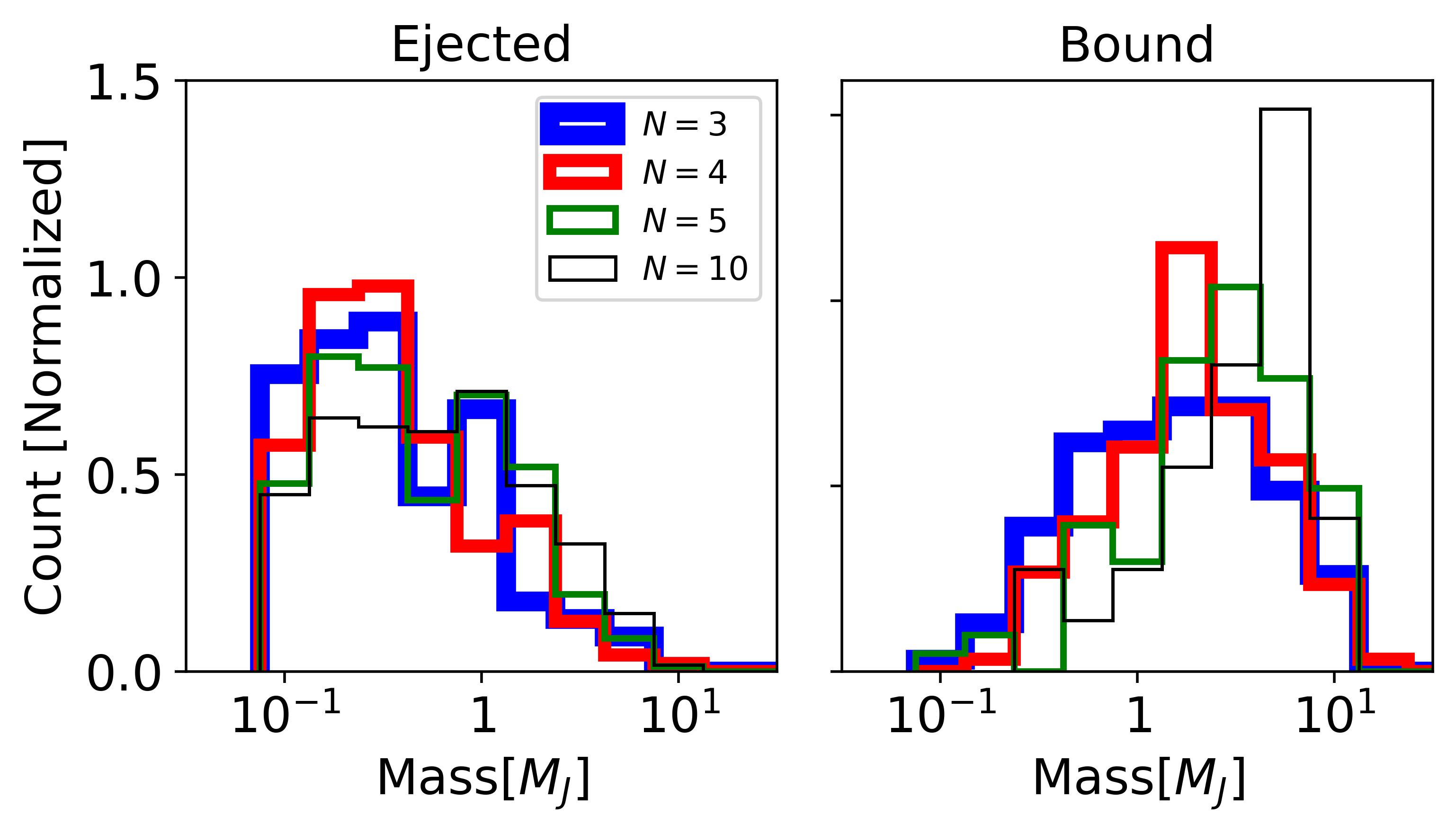}
 \caption{The mass distribution of planets ejected (left panel) and planets which remain bound (right panel) to the host star. The colors show the number of planets initially in the system. We can see that heavier planets remain bound to host star, and the lighter planets are ejected. Some of the planets which remain bound to the host star are products of collision. We use the following parameters to run the simulations: $N=5, a_{0,init}=3$ AU, $m_s=1$ $M_\odot$ and $K$=4. The mass of the planets is sampled between 0.1 and 10 $M_{jup}$ from a log uniform distribution. We evolve 100 planetary systems in our simulations.}
 \label{fig:histmdist}
\end{figure}
The mass distribution of planets ejected from the system (left) and bound to the host star (right) after $10^9$ years of evolution is shown in Figure \ref{fig:histmdist}. We can see most of the planets ejected tend to have lower masses. Meanwhile planets which remain bound tend to have higher masses. The mass distribution does not strongly depend on the number of planets initially in the system. By the looking at the right panel we can see that the some of the planets bound to the host star are products of planet-planet collisions.

\subsection{Dependence on the radius of the planet.}

\begin{figure}
	\centering
	\includegraphics[scale=0.5]{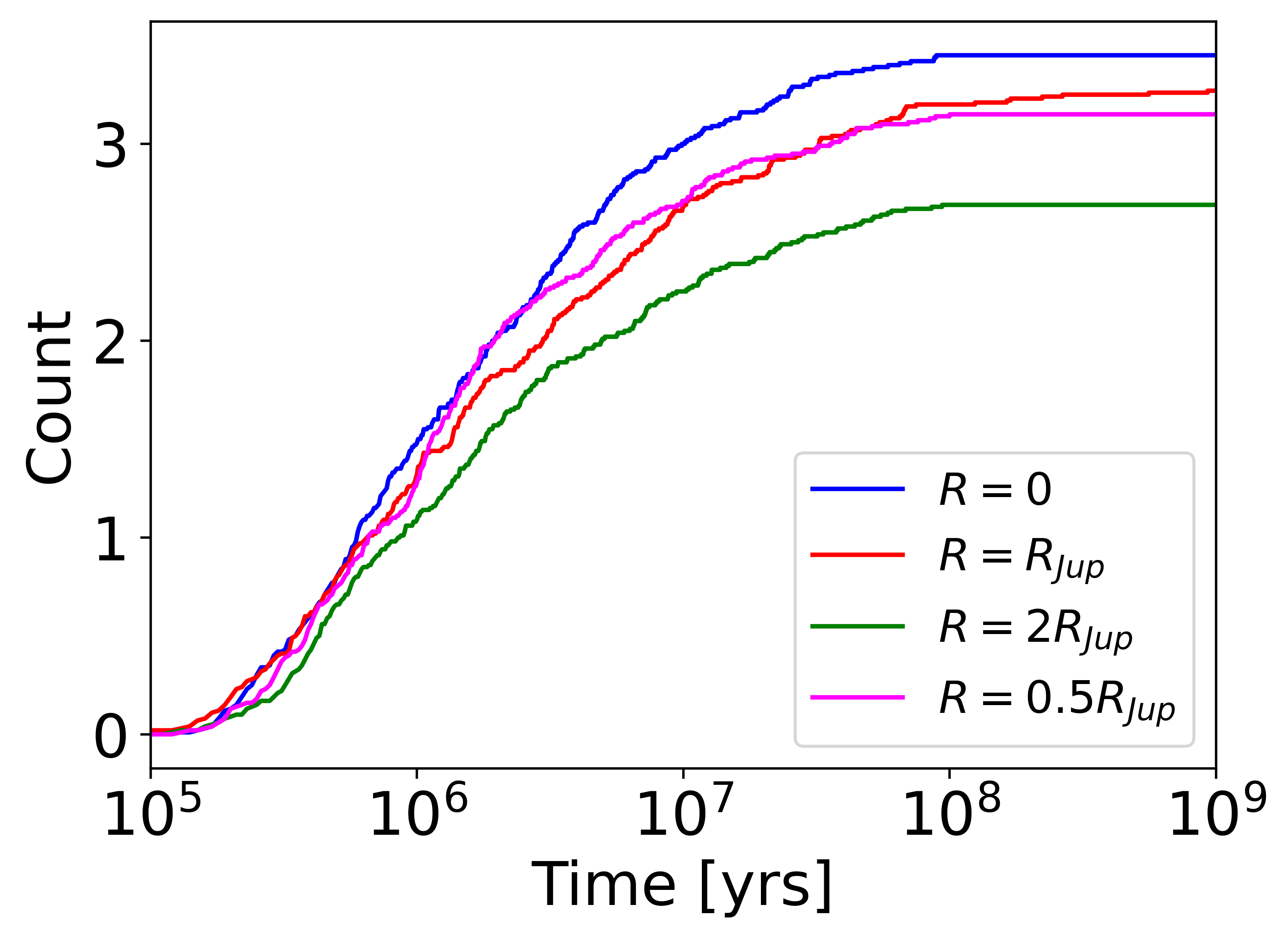}
	\caption{The average number of planets ejected from the system as a function of time. The colors show the radius of the planets. We can see that as the radius of the planets is increased, the average number of planets ejected from the system increases. The timescale over which the ejections occur does not strongly depend on the radius of the planet. We use the following parameter to run the simulations: $N=5, a_{0,init}=3$ AU, $m_s=1M_\odot,m_j=1M_{jup}$ and $K=4$. We evolve 100 systems in our simulations.}
	\label{fig:avgejprad}
\end{figure}

The mass-radius relationship of the observed population of exoplanets is well studied in literature. An empirical broken power law function is usually derived to describe the mass-radius relationship. For instance,  \cite{mullerMassradiusRelationExoplanets2024} find that the radii of smaller planets ($M<4.4M_\oplus$) are given by $R\propto M^{0.27}$. Similarly, the radii of the intermediate ($4.4M_\oplus<M<127M_\oplus$) and high mass planets ($M>127M_\oplus$) is given by $R\propto M^{0.67}$ and $\propto M^{-0.06}$ respectively. The transitions represent the change in composition of the planets. Radii of smaller and intermediate mass planets have a much steeper dependence on the mass of the planets as compared to high mass planets due to their higher densities. Also, some of the observed exoplanets have atypical radii. For instance, some of the close-in hot Jupiters have inflated radii, presumably due to higher stellar irradiation and tidal heating.  Consequently, some of the studies also include stellar flux as a parameter in their empirical fits of the mass-radius relationship. More recently, planets with even lower density (called “super puffs”) have been discovered. While most of the superpuffs are orbiting close to their host stars, the discovery of HIP 41378 f, which is on a wider orbit, has motivated the search for alternative explanations. It has been shown that rings around exoplanets can also explain their inflated raii.

In our study so far we have simulated systems in which all the planets have the mass and radius of Jupiter. It should be noted that collisions are more likely if the planets have larger radii. Collisions causes the total planetary mass to be segregated into fewer planets. Since less massive planets are usually ejected from the system, fewer planets are ejected if we increase the planetary radii.

This can be seen in Figure \ref{fig:avgejprad} which shows the average number of planets ejected from the system as function of time. Radius of the planet is shown using colors. As expected, fewer planets are ejected from the system as we increase the radius of the planets. More specifically, on average one more planet is ejected from the system in $10^9$ years if the planets were treated as point objects as compared to finite sized objects with twice the radius of Jupiter. The timescale over which ejections occur does not strongly depend on the radius of the planets. We also found that the excess velocity distribution does not strongly depend on the radii of the planets.

\subsection{Dependence on the ejection distance}

Free floating planets are defined as planets not bound to any star. In practice, due to observational constraints, stellar companions can be ruled out only within a certain distance for the planets to be labeled as ``free floating". For instance, microlensing detections of free floating planets reported by \cite{sumi_unbound_2011} could only rule out stellar companions within 10 AU. More recent detections by the OGLE survey could rule out a stellar companion within only a few AU \citep{mrozTwoNewFreefloating2019}. Similar caveats exists for photometric detections as well. \cite{miret-roigRichPopulationFreefloating2021} detect free floating planets in the Upper scorpius young stellar associations at a very low spatial density (0.4-1 FFP per square degree), strongly indicating that most of these planets do not have companions. But, \cite{bouyInfraredSpectroscopyFreefloating2022a} find a visual pair at a projected separation of 17400 AU. The authors comment that future observations of proper motions of these objects are needed to confirm if they are indeed bound. In general, widely separated pairs are common in young stellar associations (\cite{caballeroBrightestStarsOrionis2007}, \cite{joncourMultiplicityClusteringTaurus2017a}). Hence, to account for these uncertainties, we reevaluate the properties of FFPs by changing the distance from the host star beyond which we consider them to be free floating.

\begin{figure}
	\includegraphics[scale=0.5]{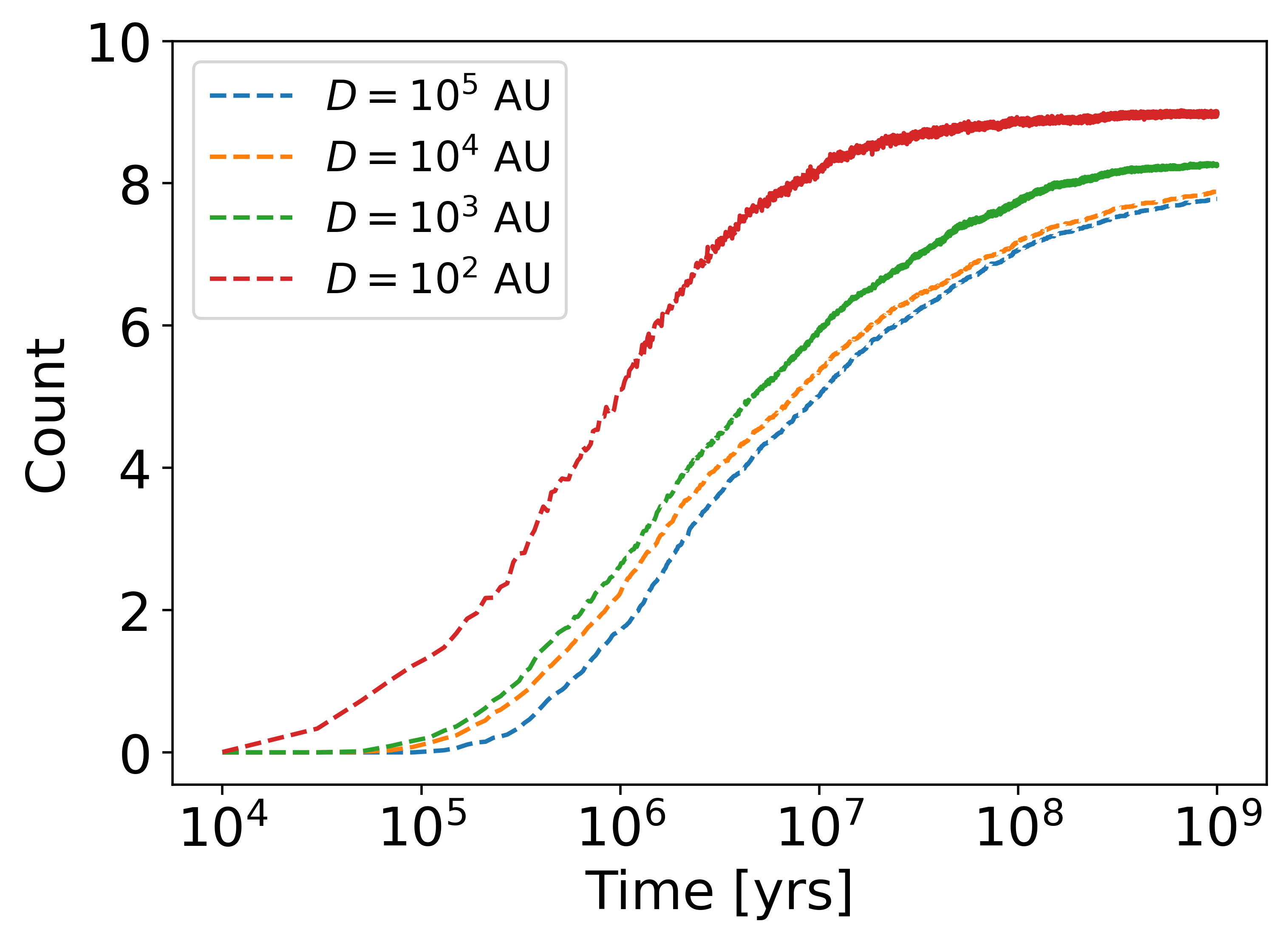}
	\caption{{ The number of planets ejected from the system as a function of time. The colors show the ejection distance beyond which the planets are considered ejected. It should be noted that we remove the planet from the system only if it reaches a distance of $10^5$ AU. We can see that the timescale over which the ejection count reaches a steady state decreases with the ejection distance. Also, on average one more planet is ejected from the system when $D=100$ AU as compared to when $D=10^5$ AU. We use the following initial conditions to make this plot: $a_{0,init}=3$ AU, $m_s=1M_\odot$, $m_i=1$ $M_{jup}$, $N=10$ and $K=4$.}}
	\label{fig:brejectnrep}
\end{figure}
{ Figure \ref{fig:brejectnrep} shows the number of planets ejected from the system as a function of time. The colors show the ejection distance. It should be noted that we remove planets from the simulations only when they reach a distance of $10^5$ AU from the host star. Consequently, many of the planets counted as ejected in Figure \ref{fig:brejectnrep} are still bound to the host star. We can see that the timescale over which the ejection count reaches steady state decreases with the ejection distance. For instance, the ejection count ($\sim 9$) reaches steady state in $10^8$ years when $D=100$ AU. On the other hand, the ejection count for $D=10^5$ AU did not attain a steady state value in $10^9$ years. On average one more planet would be considered ejected if the ejection distance is set at $10^2$ AU, as compared to $10^5$ AU. In addition, the ejection count at $10^9$ years does not vary when the ejection distance is set beyond $10^4$ AU.}

\begin{figure}
	\includegraphics[scale=0.6]{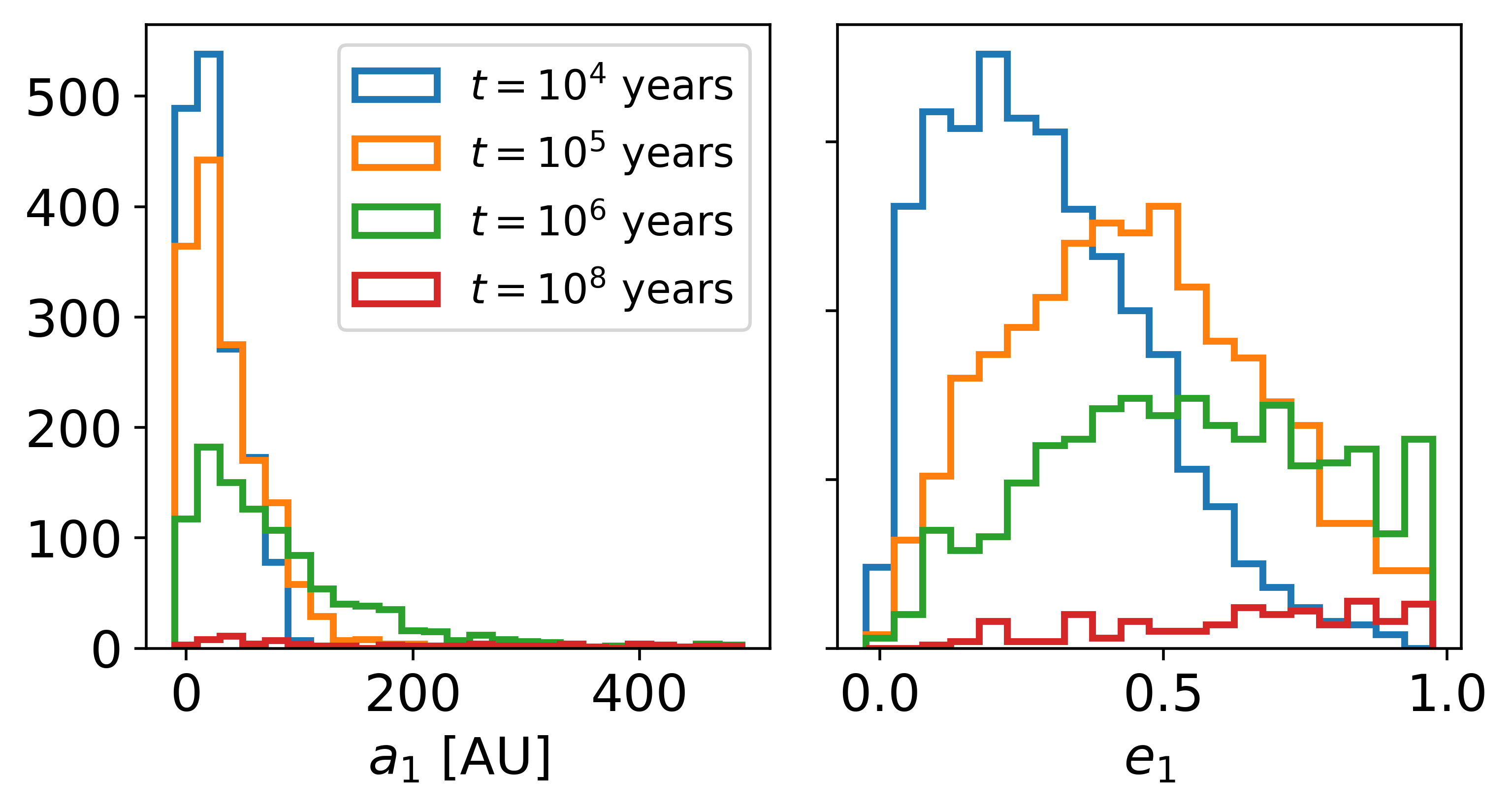}
	\caption{{  The distribution of semi-major axes (left panel) and eccentricities (right panel) of planets which are eventually ejected from the system. The colors shows the time at which the distribution was calculated. We can see that the ejected planets initially have close-in orbits ($a<100$ AU) with low eccentricities. Over time the semi-major axis distribution becomes broader, and the planets also become more eccentric. Within $10^8$ years most of the planets are ejected from the system. We use the following parameters to run the simulations: $N_p=10, a_{0,init}=3$ AU, $m_s=1$ $M_\odot,m_j=1$ $M_{jup}$ and $K$=4. We evolve 100 planetary systems in our simulations.}}
	\label{fig:histaeeject}
\end{figure}

{  Multiple widely separated planets ($a > 100$ AU) have been discovered through direct imaging. It has been suggested that these wide orbits planets could have formed through gravitational instability (e.g., \cite{bossFormationGiantPlanets2011}) or though capture (e.g., \cite{peretsOriginPlanetsVery2012,roznerBornBeWide2023}). It is also possible for the planets to be scattered into these wide orbits. Since most of the planets with host-star separation greater than 100 AU are likely to be ejected, it would be interesting to analyze the distribution of the semi-major axes and eccentricities of ejected planets as a function of time. The distribution is shown in Figure \ref{fig:histaeeject}. We can see that the initial distribution ($t<10^4$ years) of semi-major axis is confined to less than 100 AU. Initially the eccentricities of the planets are low. Over time we can see that the semi-major axis distribution becomes broader, and the planets become more eccentric. Also, the number of planets bound to the system decreases. In $10^8$ years, most of the planets are ejected from the system. The planets which do remain tend to have large eccentricities. Hence, we can conclude that wide orbit planets ($a> 100$ AU) produced by planet-planet scattering tend to have eccentric orbits, and are likely to be ejected from the system in $10^8$ years.}

\begin{figure}
	\centering
	\includegraphics[scale=0.5]{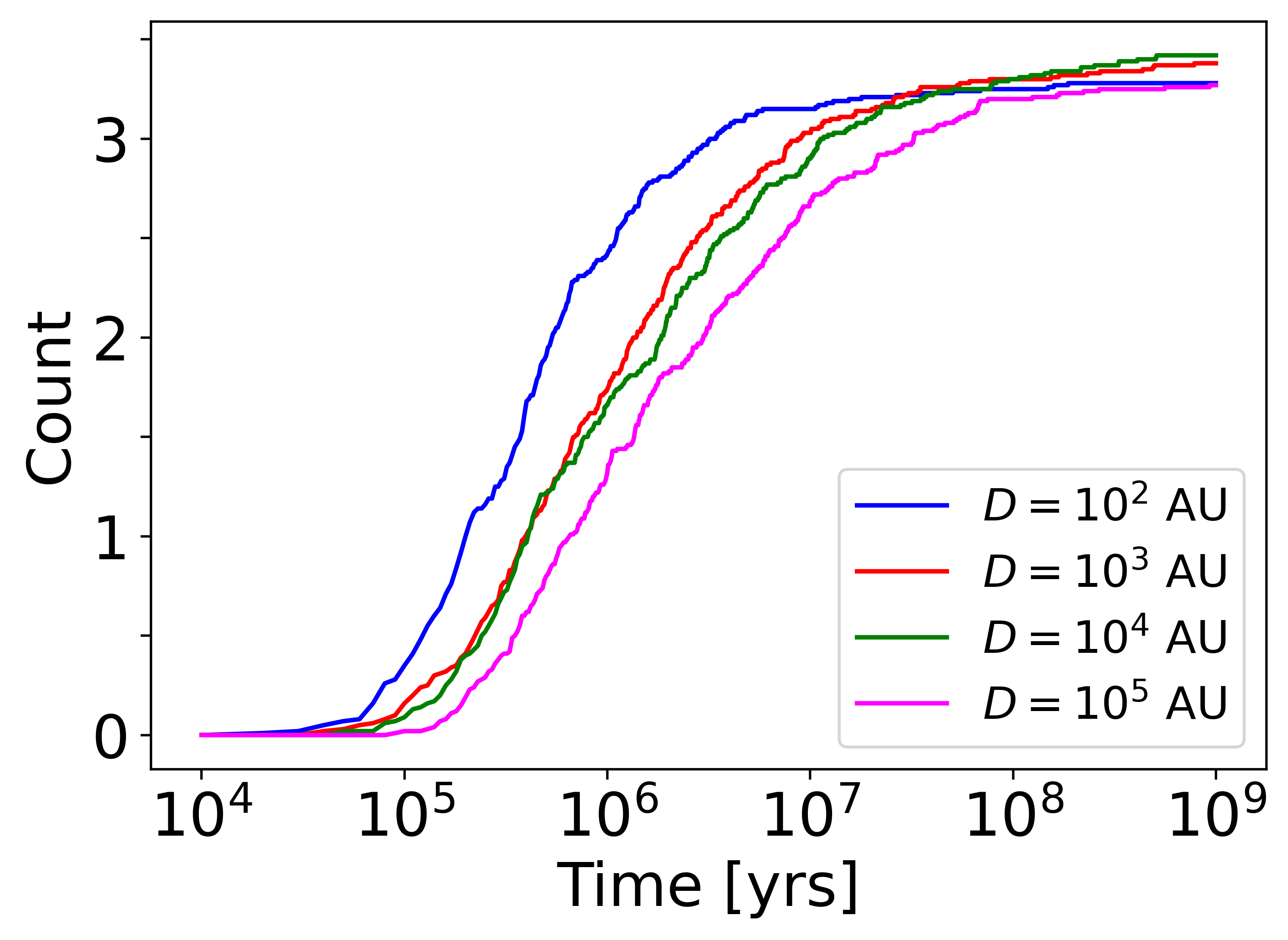}
	\caption{The average number of planets ejected from the system as a function of time. The colors show the ejection distance. We can see that total number of ejected planets does not strongly depend on the ejection distance. Also, when the ejection distance is small ($10^2$ AU), most of the planets are ejected in $10^6$ years. Meanwhile, it takes $\sim 10^8$ years for most of the planets to be ejected when $10^{3-5}$ AU. We use the following parameters to run the simulations: $N_p=5, a_{0,init}=3$ AU, $m_s=1M_\odot,m_j=1M_{jup}$ and $K$=4. We evolve 100 planetary systems for each choice of $D$.}
	\label{fig:breject}
\end{figure}

{  It should be noted that local stellar environment also dictates the outer edge of the planetary system. In our fiducial simulations we assume that planets which reach a distance of $10^5$ AU from the host star are ejected from the system. This is a natural choice in the solar neighborhood because the galactic tide becomes important beyond $10^5$ AU (e.g., \cite{heislerInfluenceGalacticTidal1986}). The galatic tide can increase the pericenter distance of a widely separated bound planet which would decouple it with the inner planetary system. In a more dense environment, like in a star cluster, the planets can be decoupled from the host exosystem at smaller distances from the host star. Stellar flybys, and stronger cluster potentials can aid in the production of FFPs. Changing the ejection distance in our simulations would also help us analyze FFP population in dense environments.}

We ran an additional set of simulations in which we vary the distance from the host star beyond which we the planet would be ejected. The ejection distance is varied between $10^2-10^5$ AU. We now {\emph{remove}} the planets from the simulation when they reach the said ejection distance. Figure \ref{fig:breject} shows the fraction of planets which are ejected from the system as a function of time. Colors show the ejection distance. As expected, the timescale over which most of the planets are ejected increases with the ejection distance. Also, the number of planets ejected in $10^9$ years does not strongly depend on the ejection distance. It should be noted that all the planets initially have a semi-major axis between 3 and 20 AU. Hence, if the semi-major axis of the planet is sufficiently excited ($a>100$ AU) due to planet-planet scattering, they are likely to be ejected from the system.
\subsection{Dependence on the initial separation between the planets}
In this study so far we have kept the initial separation between the planets constant at $K=4$ mutual Hill radii from each other. It should be noted that the initial mutual separation is not well constrained from observations. In this subsection, we change the initial separation between the planets. Within a given ensemble, we still initially place neighboring planets at constant mutual radii from each other given by parameter $K$. Between different ensembles we change the value of $K$. We simulated ensembles with $K$=2, 3, 4 and 5. In each ensemble we included 100 systems.

\begin{figure}
	\centering
	\includegraphics[scale=0.5]{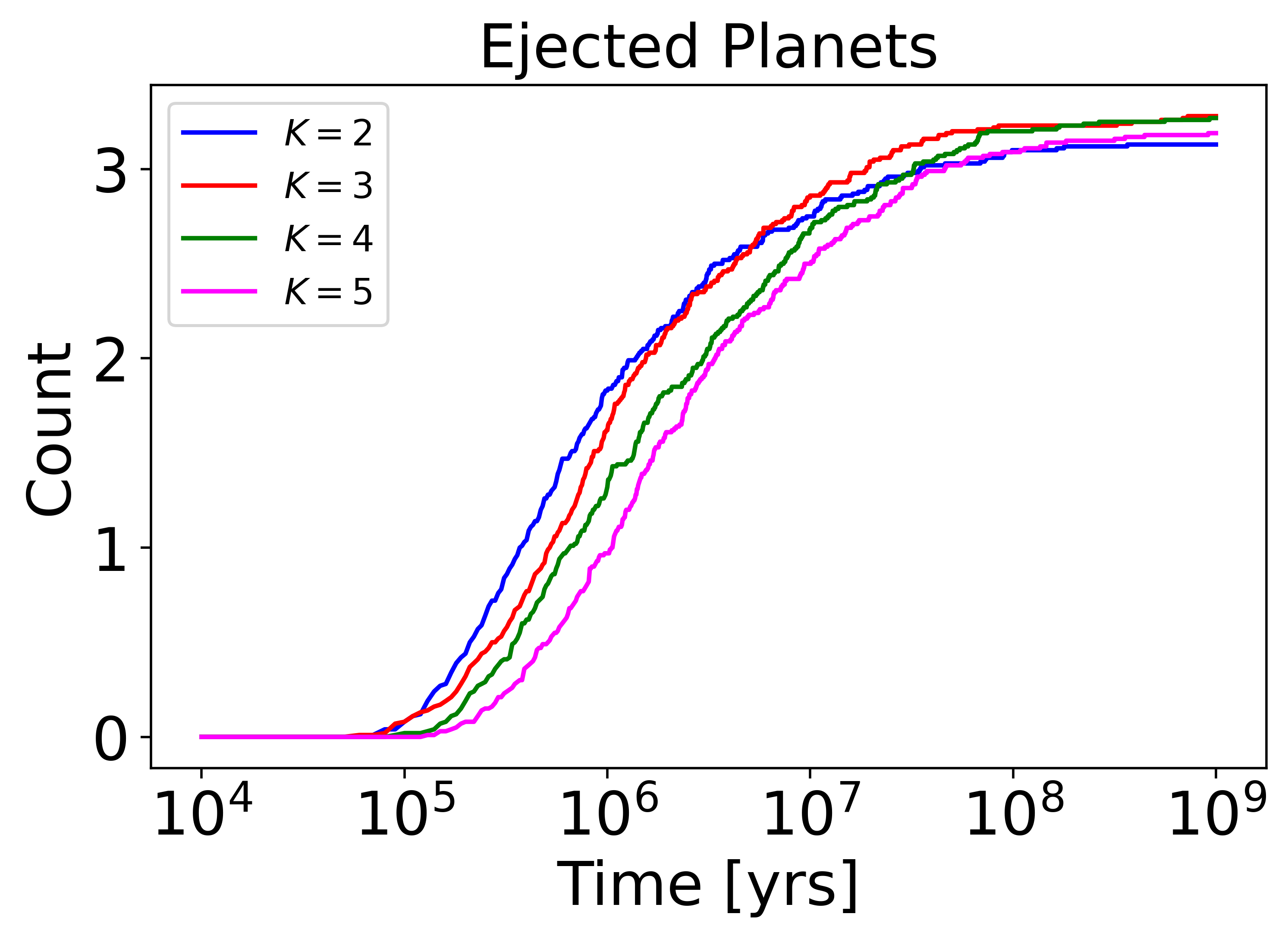}
	\caption{The average number of planets ejected from the system as a function of time. The colors show the initial separation between the planets in terms of mutual Hill radii ($K$). We can see that total number of ejected planets does not strongly depend on $K$. Also, the timescale over which the planets are ejected from the system increases with $K$. We use the following parameter to run the simulations: $N_p=5, a_{0,init}=3$ AU, $m_s=1M_\odot,m_j=1M_{jup}$ and $K=4$. We evolve 100 planetary systems in our simulations for each choice of $K$.}
	\label{fig:brejectk}
\end{figure}

Figure \ref{fig:brejectk} shows the average number of ejected planets as a function of time. The colors show the initial separation between the planets. We can see that the ejection timescale increases with $K$. For instance, on average, it takes twice as long for a planet to be ejected from the system when $K=5$, as compared to when $K=2$.  Also, for $K=5$ the ejection count does not reach steady state in $10^9$ years. This is consistent with previous empirical results which show that the instability timescale increases with $K$. We can also see that the total number of planets ejected from the system in $10^9$ years does not strongly depend on $K$. We also find that the average excess speed of the ejected planets increases as $K$ decreases: for $K=5$, the average excess speed is 3.1 km/sec, meanwhile when $K=2$, the average excess speed increases to 3.9 km/sec. This is due to the fact that as $K$ decreases, the relative velocity between the planets decreases, which allows stronger scattering encounters between the planets.
\section{Discussion}
\label{sec:discuss}
\subsection{Comparison with observations.}

\begin{figure}
	\centering
	\includegraphics[scale=0.5]{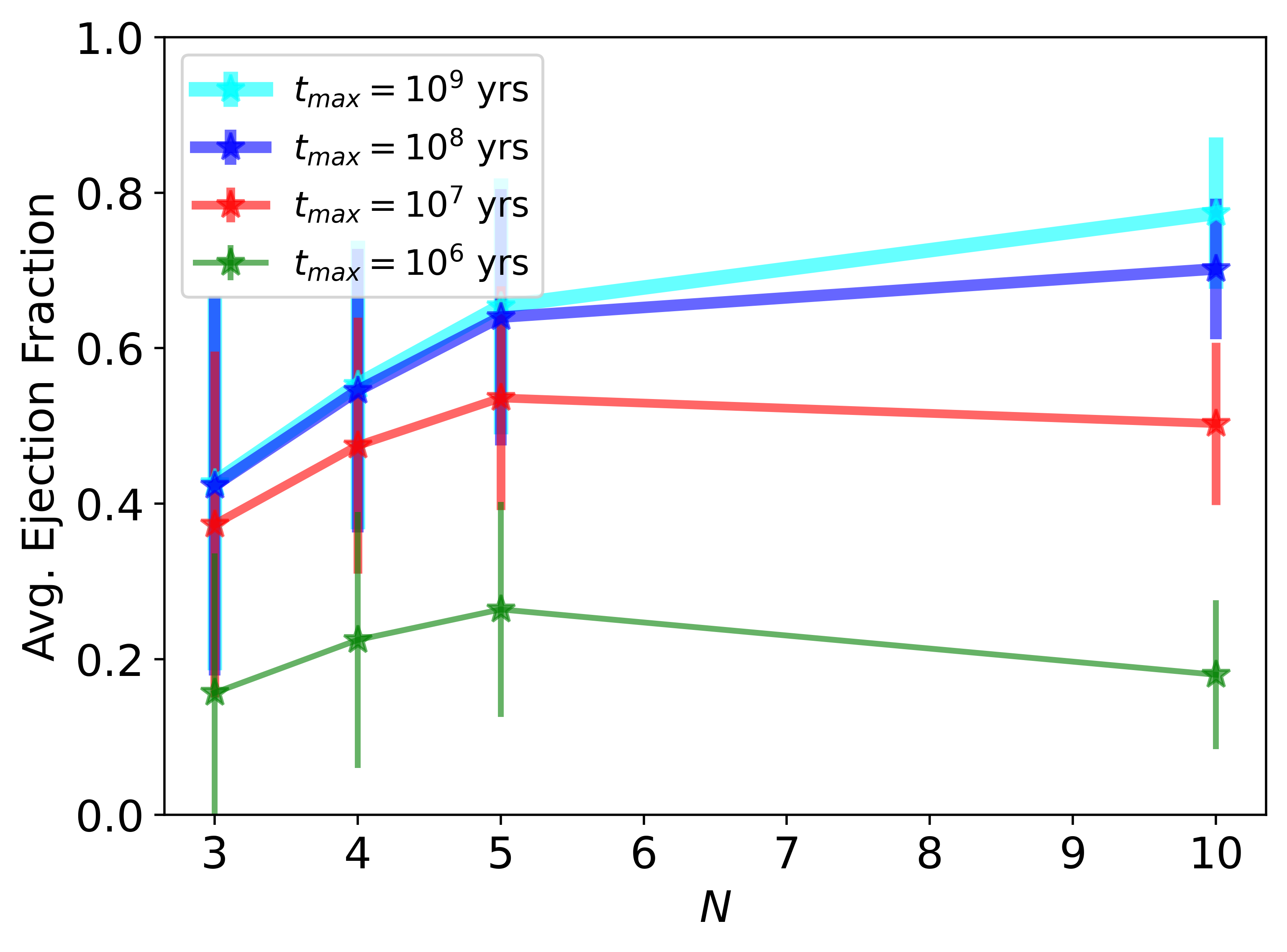}	
	\caption{The average fraction of planets which are ejected from the system at different times in the simulation. The x-axis shows the number of planets initially in the system, and the y-axis shows the fraction of planets ejected from the system. The colors indicate the time at which the numbers are recorded. We can see that, in general, the ejection fraction increases with time from around $20\%$ at $10^6$ years, to more than $60\%$ at $10^8$ years.}
	\label{fig:ejectfractavg}
\end{figure}

\begin{figure}
	\centering
	\includegraphics[scale=0.5]{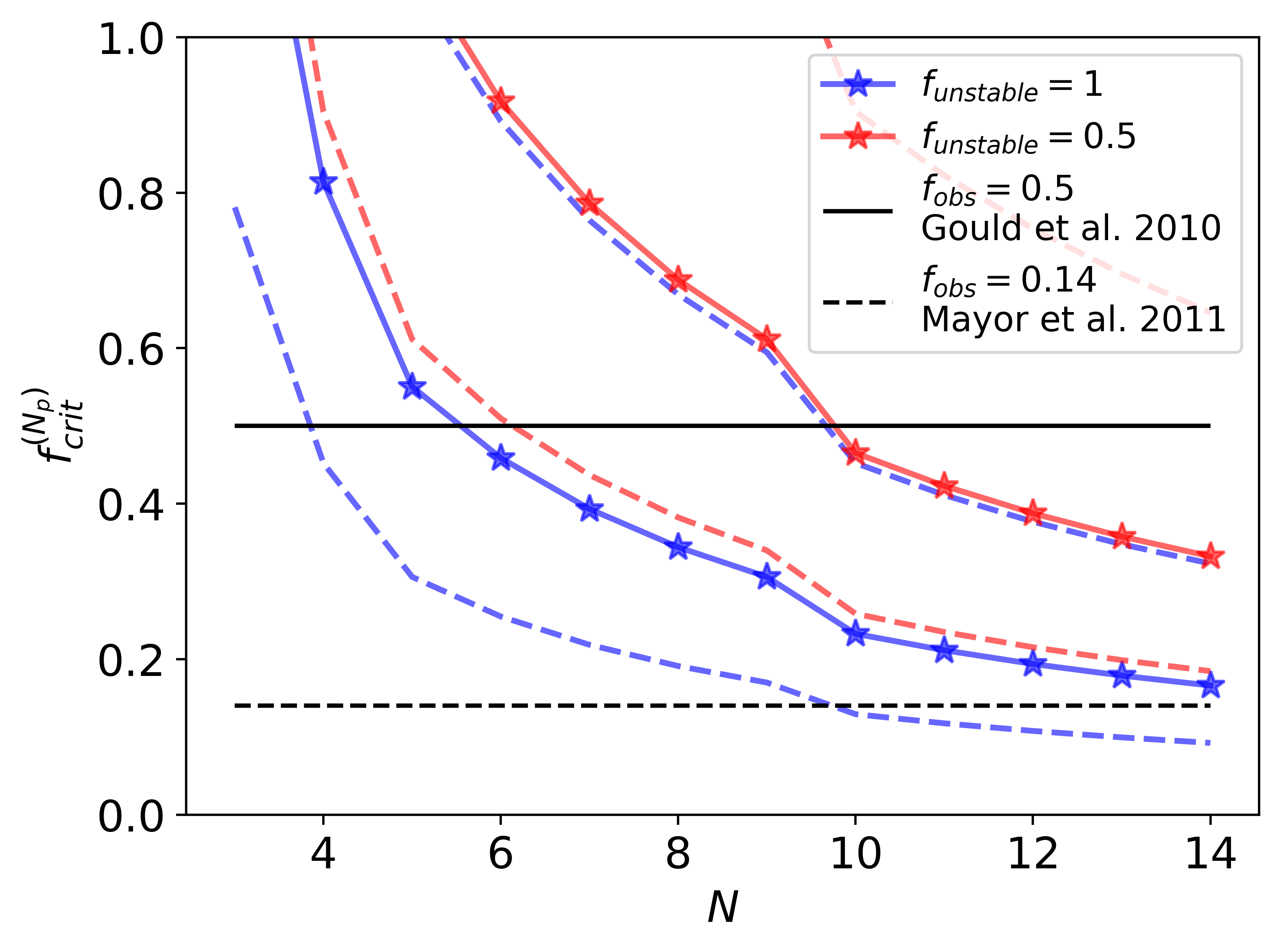}	
	\caption{The critical fraction of stellar systems that must contain $N$ planets to produce the observed number of free floating planets. The limits from observations are shown using the horizontal black lines. The results shown using the blue (red) line assume that all ($50\%$) of the exo-systems are unstable.}
	\label{fig:critnp}
\end{figure}
We will now compare the number of planets ejected from the planetary system through planet-planet scattering with the observed census of free floating planets. We will repeat the analysis done in \cite{veras_planet-planet_2012}, with  our N-body results. The number of FFPs generated through planet-planet scattering depends on the number of planets which form in the system, the fraction of planetary systems which are unstable, and the ejection fraction.  The number of FFPs per star which are ejected from the system is hence given by:
\begin{equation}
	\frac{N_{free}}{N_{stars}} = \sum_{N_p=3}^{\infty} f_{giant}^{N_p} f_{unstable}^{N_p} f_{eject}^{N_p} N_p
\end{equation}
where $f_{giant}^{N_p}$ is the fraction of exosystems with $N_p$ giant planets, $f_{unstable}^{N_p}$ is the fraction of the systems which are unstable, and $f_{eject}^{N_p}$ is the ejection fraction. 

The ejection fraction ($f_{eject}^{N_p}$) can be deduced from our N-body simulations. Figure \ref{fig:ejectfractavg} shows the average ejection fraction from all of our simulations as a function of time. We can see that the ejection fraction can significantly increase with time. For instance, for $N_p=10$ the ejection fraction increases from $\sim 20\%$ in $10^6$ years to $\sim 80\%$ in $10^9$ years.  Also, the ejection fraction increases with the number of planets initially in the system. In general, $40-80 \%$  of the planets are ejected from the system. 

The fraction of exosystems with $N_p$ planets ($f^{N_p}_{giant}$) is constrained by observations. Multiple estimates are available in literature. For instance, \cite{gouldFREQUENCYSOLARLIKESYSTEMS2010} finds that half of all stars have planetary systems. Hence, $\sum_{N_p=3}^{\infty}f_{giant}^{N_p} = f_{giant}^{total} < 0.5$. Meanwhile \cite{mayorHARPSSearchSouthern2011} find $f_{giant}^{total} > 0.14$. 

A critical fraction of systems ($f_{crit}^{N_p}$) which should have $N_p$ planets to produce the observed number of FFPs can be obtained by the following expression:
\begin{equation}
	f_{giant,crit}^{N_p} = \left(\frac{N_{free}}{N_{stars}}\right)_{obs} \frac{1}{N_{p}f_{unstable}^{N_p}f_{eject}^{N_p}}
\end{equation}
It should be noted that the number of exosystems which are unstable is unconstrained. In our analysis we chose two values of $f^{N_p}_{unstable}=0.5$ and 1. Smaller number of planets are needed to explain FFP observations if more exosystems are initially unstable. 

We show the critical giant planet fraction from our simulations in Figure \ref{fig:critnp}. The results shown in blue correspond to $f^{N_p}_{unstable}=1$, and the results shown in red correspond to $f^{N_p}_{unstable}=0.5$. The solid curve corresponds to the nominal of $(N_{free}/N_{stars})_{obs}=1.8$  (from \cite{sumi_unbound_2011}) and dashed curves show the observational uncertainties.  The horizontal black lines show the observations constrains on $f^{total}_{giant}$. It can be seen that our results are shifted to the left as compared to the results of \cite{veras_planet-planet_2012}. This is due to the fact that we run our simulations for longer timescales, and hence our ejection fractions are higher. Consequently, we find that fewer planets need to form in exosystems to explain the observed population of FFPs. In general, atleast 4 planets should form in the exosystems to explain the observation. To produce the nominal population of the FFPs, 5-10 planets should form in the exosystems. Despite being high, the current planet formation models allow these number of planets to be formed around stars.

\subsection{Properties of the bound planets}

\begin{figure*}
	\centering
	\includegraphics[scale=0.9]{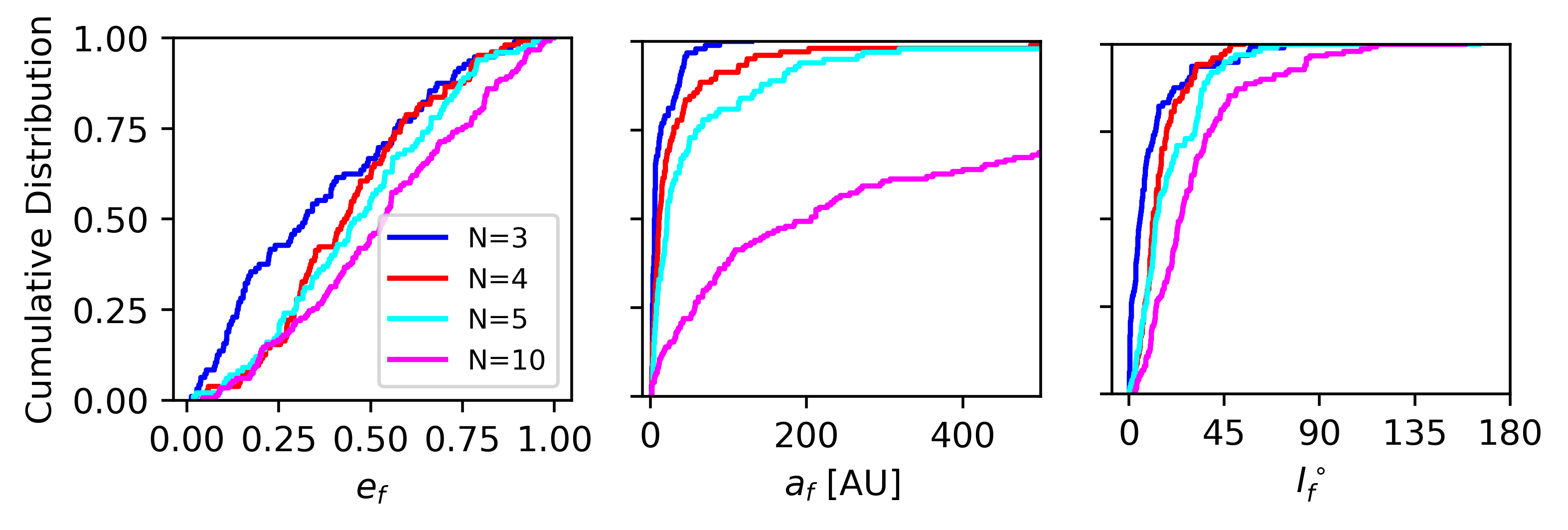}	
	\caption{The figure shows the cumulative distribution of eccentricity (left panel), semi-major axis (middle panel), and the inclination (right panel) of bound planets at the end of the simulation. The colors shows the initial number of planets in the system. We use the following initial conditions to make this plot: $a_{0,init}=3$ AU, $m_s=1M_\odot$ $m_i=1$$M_{jup}$, and $K=4$.}
	\label{fig:distbound}
\end{figure*}

We will now look at the properties of planets which remain bound to the star at the end of the simulation. It should be noted that the number of bound planets at the end of the simulation depends on $N$. For $N=3$, most of the systems have 1 bound planet at the end of simulation. Also, when $N=4,5$ most systems have 2 planets, and for $N=10$ most systems have 1-3 bound planets.

Figure \ref{fig:distbound} shows the cumulative distribution of the orbital elements of the bound planets. The left, middle and the right panels show the distribution of eccentricities, semi-major axes and inclination. The color indicates the number of planets initially in the system. We can see the distribution of eccentricities is broad, with significant number of planets on very eccentric orbits. For instance, $20\%$ of bound planets in N=3 systems have an eccentricity greater than 0.8. In general, the eccentricity of bound planets increases with the number of planets initially in the system. In the simulations shown here, the mean eccentricity increases from $0.43$ for $N=3$ to $0.63$ for $N=10$.  Also, we do not find a strong corelation between initial location of the planet, and it's final eccentricity.

\begin{figure*}
	\includegraphics{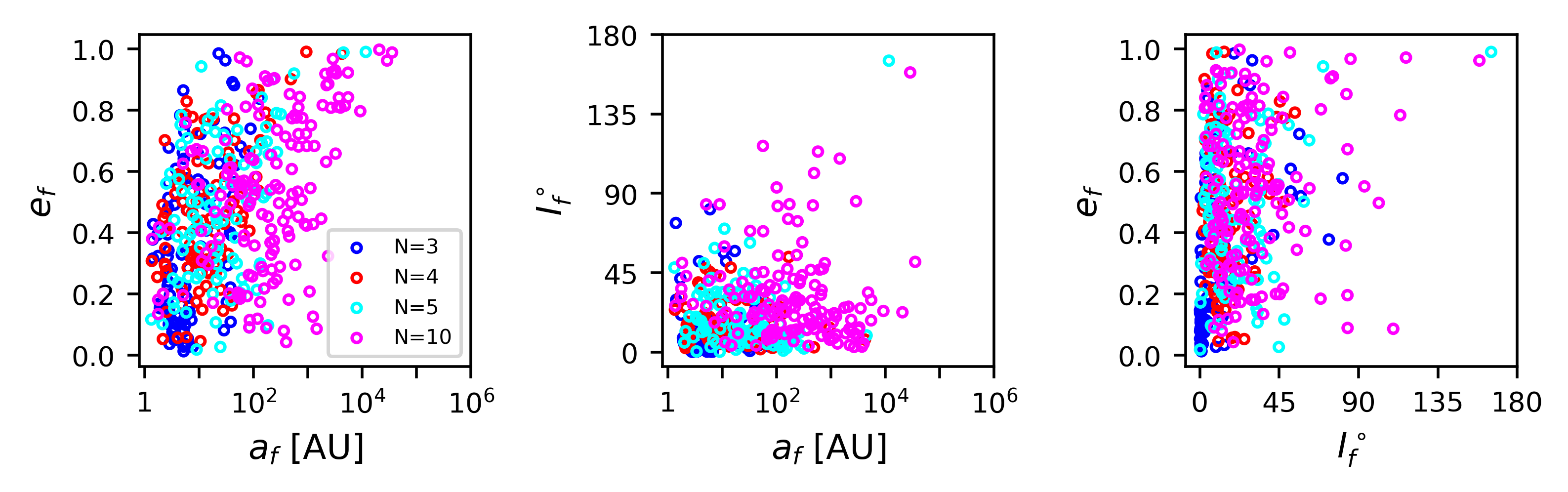}
	\caption{Orbital properties of planets which remain bound to the host star at the end of the simulation. The left panel shows semi-major axes (x-axis) vs eccentricities (y-axis), the middle panel shows the inclination (y-axis) vs semi-major axis (x-axis) and the right panel shows the eccentricities (y-axis) vs inclinations (x-axis) of the bound planets. The colors show the number of planets initially in the system. We use the following initial conditions to make this plot: $a_{0,init}=3$ AU, $m_s=1M_\odot$ $m_i=1$ $M_{jup}$, $R_i=1$ $R_{jup}$, and $K=4$.}
	\label{fig:scatteraei}
\end{figure*}

Planet-planet scattering can significantly alter the semi-major axes of planets. As compared to the initially compact planetary systems, the final distribution of the semi-major axes of bound planets is broad. The median semi-major axis for $N$=3, 4, 5 and 10 are $7.2,$ $17.5,$ $32.2$ and 209 AU respectively. As expected, the distribution is broadest for $N=10$ systems. We find that the final semi-major axis of bound planets depends on their initial semi-major axis. More specifically, planets which have wide final orbits usually are initialized on wide orbits. The systems with more planets tend to more planets farther away from the host star, where the potential well of the host star is shallow. As a result, scattering is more effective in exciting the semi-major axis of the planets when $N$ is larger. { In addition to scattering it should be noted that wide planet systems can also originate from capture \citep{peretsOriginPlanetsVery2012,roznerBornBeWide2023}. A comparison of the populations is left to a future study.}

The rightmost panel of Figure \ref{fig:distbound} shows that the inclination of bound planets can be excited to large values. For instance, a significant fraction of planets ($7-33\%$) can end up with an inclination $>40^\circ$. The inclination distribution is broader for systems with larger $N$. The final inclination does not have a strong dependence on initial location of the Planet.

Figure \ref{fig:scatteraei} shows the orbital elements of planets bound to the host star. We can see that planets on wide orbits tend to have large eccentricities. This trend seem to hold for systems with different number of planets initially in the system.  Also, retrograde planets produced by planet-planet scattering tend to be on wider orbits ($a>100$ AU). In addition, retrograde orbits tend to have higher eccentricities ($e>0.5$).

In most of our fiducial simulations, despite collisions, $\sim90\%$ of bound planets had the same final mass as the initial mass of the planets. The remaining $10\%$ of planets had twice the initial mass as a result of 1 planet-planet collision. The mass distribution was different for $N=10$ systems in which $~98\%$ of bound planets had the same final and initial masses.

Our results are consistent with previous works on scattering in multi-planetary systems. For instance,  \cite{chatterjeeDynamicalOutcomesPlanetPlanet2008} study planet-planet scattering in three planet systems. Instead of assigning same masses to all the planets, they use three different mass distributions which are motivated by planet formation theories. Nevertheless, results shown in Figure \ref{fig:distbound} are in broad agreement with results of \cite{chatterjeeDynamicalOutcomesPlanetPlanet2008} (See their Figures 3,5 and 9). Similar to our work, \cite{chatterjeeDynamicalOutcomesPlanetPlanet2008} find that the median eccentricity of bound planets is around 0.4, and the median inclination is around $20^\circ$. Our distribution of final semi-major axes is broader than the distribution shown in \cite{chatterjeeDynamicalOutcomesPlanetPlanet2008}. 

\cite{petrovichSCATTERINGOUTCOMESCLOSEIN2014} study planet-planet scattering in close-in systems with two or three planets ($a<0.15$ AU).  In these close-in systems, collisions between the planets are important.	As expected, they find that ejection fraction depends strongly on the semi-major axis of the planets. They are relatively rare ($1\%$) in close-in planets ($a<0.15$ AU) as compared to widely separated planets ($\sim 30\%$) at a few AU.

\section{Conclusions}
\label{sec:conc}
In this work we run an ensemble of N-body simulations of planetary systems. We focus on the planets ejected from the system due to planet planet scattering. The planets are initialize on near circular, near coplanar orbits, similar to the expected configuration after planet formation. We run our simulations for a maximum of $10^9$ years. { It should be noted that most studies in literature focus on timescales less than $10^9$ years.} Between our simulations we vary the number of planets in the system, mass distribution of the planets, initial separation between the planets, semi-major axis of the inner most planet and the radii of the planets. 

We find that between 1 to 8 planets are ejected from the system depending on the number of planets initially in the system. Most of the ejections happen within $10^8$ years. If the planetary system initially has $\sim 10$ planets, the ejection fraction does not reach steady state even in $10^9$ years. Planets are equally likely to be ejected from the system irrespective of their initial location. Meanwhile, collisions which are more likely to happen in the inner part of the planetary system. 

The excess speed of the ejected planets has a broad distribution. The mean excess speed is generally in the range of 2-6 km/sec. Some of the ejected planets can have excess speeds greater than 20 km/sec. Our results show that the excess speed of the ejected planet strongly depends on the semi-major axis of the inner most planet. Also, the excess speed does not vary much with the number of planets in the system. We also analyzed the direction of the planets ejected from the planetary system with respect to the initial angular momentum of the planetary system. We found that most of the planets are ejected in the plane of the planetary system. The planets which are close-in to the host star are more likely to be ejected from at an inclined angle. In general, most of the planets ($> 70\%$) are ejected within $\sim 30^\circ$ of the plane of the planetary system.

Changing the initial separation between the planets does not significantly affect the properties of the ejected planets. Both the ejection fraction, and the excess velocity distribution remain unaltered. Meanwhile, the timescale over which ejections occur does increase with the initial separation between the planets. 

Increasing the radii of the planets reduces the ejection fraction of the system. This is due to the fact that collisions between the planets becomes more likely with the increased radii of the planets. We also found that the excess speed of the ejected planets does not strongly depend on planetary radii. 

{ We also ran an ensemble of simulations in which the initial masses of the planets are sampled from a log uniform distribution. We find that the mass distributions of the bound and ejected planets are distinct. The bound planets tend to have higher masses as compared to planets which are ejected from the system.}

To address the constraints on the methods used to detect free floating planets, we vary the distance from the host star beyond which the planet is considered “free floating”. We find that the ejection distance does not significantly change ejection fraction.  Meanwhile, reducing the ejection distance can reduce the timescale it takes for the ejection fraction to reach steady state. 

Finally we compare our results with the observed populations of FFPs. We find that $5-10$ planets should form per star to reproduce the observed population of FFPs. This indicates that if not whole, part of the FFP population can be explained by planet-planet scattering.
\bibliography{ffppaper,ffppaper1}{} 
\bibliographystyle{aasjournal}
\end{document}